\PassOptionsToPackage{table}{xcolor}
\documentclass[2022,pdflatex,1p,times,twocolumn,12pt]{elsarticle}

\makeatletter
\renewcommand{\fnum@figure}{Fig. \thefigure}
\makeatother
\usepackage{float}
\usepackage{cuted}
\usepackage{graphicx}
\usepackage[utf8]{inputenc}
\usepackage{pdflscape}

\usepackage[table]{xcolor}
\usepackage{array}

\usepackage{adjustbox}
\usepackage{mdframed}
\usepackage{multirow}
\usepackage{xcolor} 
\definecolor{olive}{RGB}{112,130,56}

\usepackage{longtable}
\usepackage[labelfont=bf]{caption}

\usepackage{adjustbox}
\usepackage{graphicx}

\usepackage{float}
\usepackage{multirow}
\usepackage{bm}
\usepackage{subfig}
\usepackage{array}
\arrayrulecolor[HTML]{fbebd3}

\usepackage{float}
\usepackage[utf8]{inputenc}
\usepackage{pdflscape}

\usepackage[table]{xcolor}
\usepackage{array}

\usepackage{adjustbox}
\usepackage{mdframed}
\usepackage{multirow}
\usepackage{xcolor} 
\definecolor{olive}{RGB}{112,130,56}

\usepackage{longtable}

\usepackage{adjustbox}
\usepackage{graphicx}
\usepackage{float}
\usepackage{multirow}
\usepackage{bm}
\usepackage{subfig}
\usepackage{array}
\arrayrulecolor[HTML]{fbebd3}

\usepackage{amssymb}
\usepackage[colorlinks=true,linkcolor=black,anchorcolor=black,citecolor=black,filecolor=black,menucolor=black,runcolor=black,urlcolor=black]{hyperref}

\begin{document}

\begin{frontmatter}

\title{\textsf{Nuclei \& Glands Instance Segmentation in Histology Images : A Narrative Review}}


\author[1]{Esha Sadia Nasir}

\affiliation[1]{organization={School of Electrical Engineering \& Computer Science (SEECS)},
            addressline={National University of Sciences and Technology (NUST)}, 
            city={Islamabad},
            country={Pakistan}}

\author[1]{Arshi Parvaiz}
\author[1]{Muhammad Moazam Fraz}

\begin{abstract}
Instance segmentation of nuclei and glands in the histology images is an important step in computational pathology workflow for cancer diagnosis, treatment planning and survival analysis. With the advent of modern hardware, the recent availability of large-scale quality public datasets and the community organized grand challenges have seen a surge in automated methods focusing on domain specific challenges, which is pivotal for technology advancements and clinical translation. In this survey, 126 papers illustrating the AI based methods for nuclei and glands instance segmentation published in the last five years (2017-2022) are deeply analyzed, the limitations of current approaches and the open challenges are discussed. Moreover, the potential future research direction is presented and the contribution of state-of-the-art methods is summarized. Further, a generalized summary of publicly available datasets and a detailed insights on the grand challenges illustrating the top performing methods specific to each challenge is also provided. Besides, we intended to give the reader current state of existing research and pointers to the future directions in developing methods that can be used in clinical practice enabling improved diagnosis, grading, prognosis, and treatment planning of cancer. To the best of our knowledge, no previous work has reviewed the instance segmentation in histology images focusing towards this direction.
\end{abstract}




\end{frontmatter}


\section{Introduction}\label{sec1}

\begin{figure}[H]
\centering
\includegraphics[width=1\textwidth, height = 4cm]{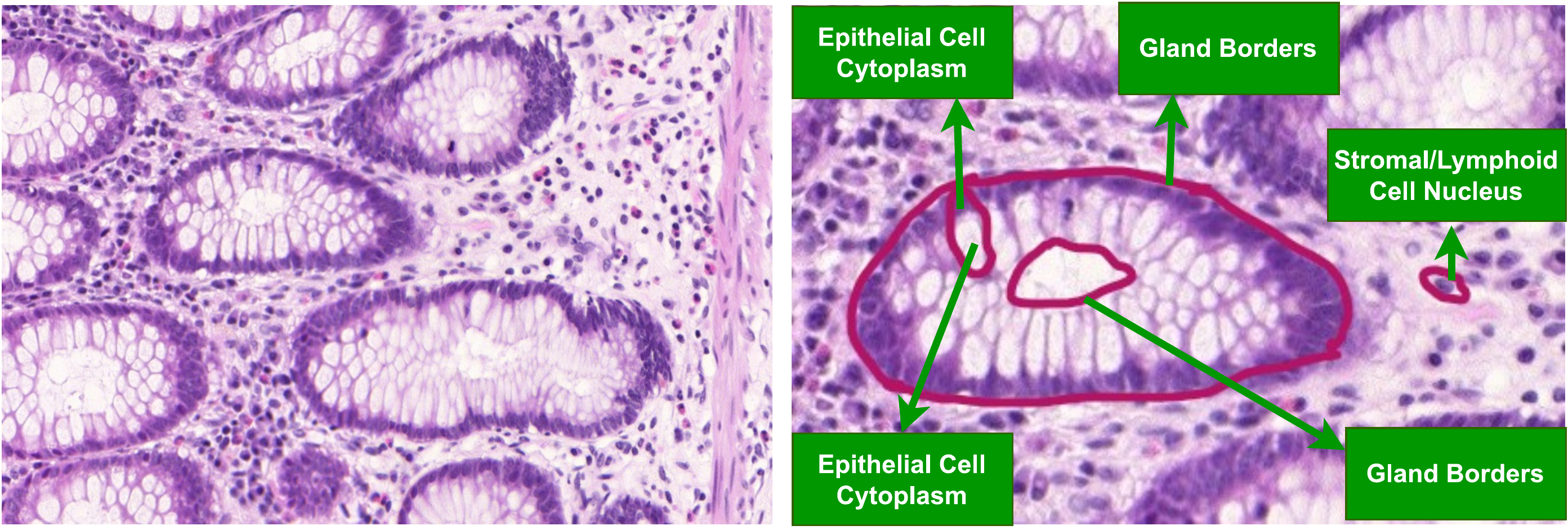}
\caption{left: An image of a colon tissue, right: an individual gland colon tissue with sub structures. }\label{fig1}
\end{figure}

In the last few decades, the advent of computational pathology  has catalyzed the advancements in clinical diagnosis, expedited development of new interactive models for pathology education and paved way for incredible rise in whole slide image analysis tools. It has revolutionized the entire tissue specimen analysis process for pathologists. From manually analyzing thousands of tissue slides via microscope requiring specialized doctors to automatic digital slide generation via scanning  and using  AI based deep learning techniques has  spawned  fatal disease diagnosis such as cancer using image analysis. 
In digital pathology nuclei and gland instance segmentation in whole slide images is of pivotal value for abnormality assessment. It plays a key role in histopathological image analyses  whether it be identification of  major chronic disease including tumor localization through segmentation or classification as benign or malignant.
Glands are often considered as one of the main histological structures present in most of the organs as primary mechanism for proteins and carbohydrates secretion. However, it has been observed that adenocarcinomas, regarded as the most severe type of cancer, originates from glandular epithelium as malignant tumors. In figure \ref{fig1}, left patch shows histopathological image of a colon tissue stained with routinely used Haematoxylin and Eosin technique  while right patch is individual gland of a colon tissue with sub structures.
This makes understanding of glands morphology a pivotal step for assigning degree of malignancy of major adenocarcinomas e.g in breast, colon, lung and prostate. Thus accurate gland instance segmentation is considered as a necessary step for obtaining valid morphology information.

Accurate segmentation and classification is of crucial importance during examination of crisscrossing cellular events in the process of wound healing by Oswal et al. \cite{5678}, cancer grading which is grade assignment on the basis of trends observation of molecular and morphological changes i.e variation in shapes, size, texture and orientation as well as other clinical assessments in cell nuclei quantification which is of strong prognostic significance during bio marker determination for immune tumor infiltrate quantification and other various pathological assessments is studied by Jung et al. \cite{Jung2019AnAN} and Benjamin et al.\cite {count, 9012}.
In comparison to various other medical fields including neuroscience, cardiology, radiology, ophthalmology and dermatology. Cell Microscopy and Computational Pathology as referenced by Ahmed et al. \cite{1234} is the most prominent application of neural networks because of expeditious transition of classical histopathology to computational pathology has revolutionized the entire image analysis workflow. From manually reading and annotating thousands of slides to digitizing biopsy process resulted in an increased demand for predictive analysis that enables the election and stratification of patients for further treatment \cite{application}.
Medical Imaging competitions have been an effective approach to crowd source
the development of  new techniques  and are a major reason behind increase research interest towards whole slide image analysis because of the introduction of grand challenges after every new  public dataset launch including NuCLS \cite{Amgad2021NuCLSAS}, BACH \cite{8012}, MoNuSeg \cite{8880654} and CoNIC \cite{7012}. 


Various medical image analysis studies regards segmentation of cell nuclei as a preliminary step for extracting meaning biological insights. In research studies including protein localization, moving population track, phenotype classification, profiling treatments and many more are considered as a reliable candidate for identification of single cells microscopy images. It also serves as a prerequisite for CAD systems in computational pathology, where features such as nuclear pleomorphism and cytology of the nucleus can assist in making a prognosis as discussed by Graham et al. \cite{8363645}.

\begin{figure}[H]
    \centering
    \begin{tabular}{cccc}
        \subfloat[]{\includegraphics[width=3.1cm,height=3.1cm]{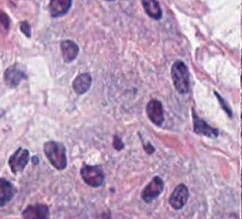}} &
        \subfloat[]{\includegraphics[width=3.1cm,height=3.1cm]{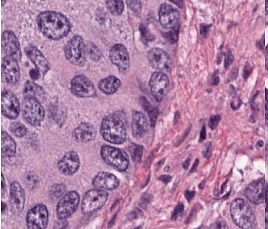}} &
        \subfloat[]{\includegraphics[width=3.1cm,height=3.1cm]{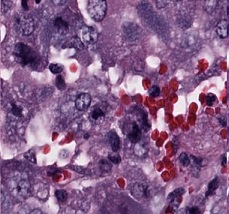}} &
        \subfloat[]{\includegraphics[width=3.1cm,height=3.1cm]{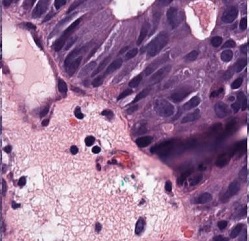}} \\

        \subfloat[]{\includegraphics[width=3.1cm,height=3.1cm]{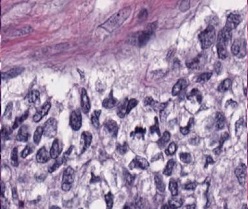}} &
        \subfloat[]{\includegraphics[width=3.1cm,height=3.1cm]{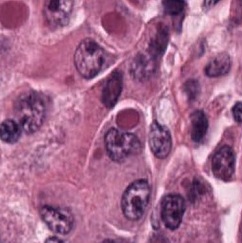}} &
        \subfloat[]{\includegraphics[width=3.1cm,height=3.1cm]{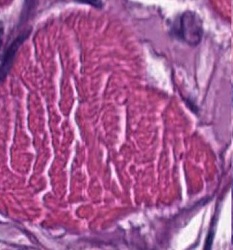}} &
        \subfloat[]{\includegraphics[width=3.1cm,height=3.1cm]{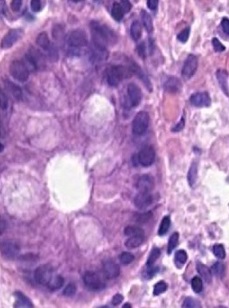}} \\
    \end{tabular}
    \caption{A catalogue of morphological variation in  nuclei at 20x magnification for different organs.  \textbf{(a)} Pancreas \textbf{(b)} Esophagus \textbf{(c)} Adrenal Gland \textbf{(d)} Cervix \textbf{(e)} Bile duct \textbf{(f)} Liver \textbf{(g)} Oral \textbf{(h)} Colon}
    \label{fig2}
\end{figure}

However, nuclei and glands instance segmentation cannot be regarded as a simple task for non experts in pathology labs. Despite being continuous advancements in learning algorithm, nuclei segmentation is still an extremely challenging task because of blurred nuclei boundaries, differences in size and shape highlighted by Vahadane et al. \cite{inproceedings}, uneven staining, pathological changes on pathological images,  morphological abnormalities \cite{Sirinukunwattana2016LocalitySD} and substantial color variations described by Rashmi et al. \cite{colornorm}. Similarly, varying morphology of glands at different histological grades, different intrinsic features of glands WSIs poses major challenge during segmentation of instances. Firstly, applying mathematical shape model for instance segmentation gets difficult due to shapes heterogeneity. Figure \ref{fig2} shows structural variations of nuclei in different organs. Secondly granule filled  cytoplasm cause nucleus extrusion to flat shapes as compared to  oval or round structures in normal cases mentioned by Yan et al. \cite{gland2}. Thirdly, cellular matrix variations results in anisochromasia thus resulting in additive noise in background compared to normal intensity gradients.

Improper staining often times result in similarity in nuclei  to cytoplasm or background  colors thus yielding blurred boundaries \cite{9656603}. Moreover occurrence of several overlapping nuclei in whole slide images causes further  difficulty in objects segmentation. 
Major challenge in  model development is varying types of nuclei e.g nerve cell nuclei are typically triangular in shape \cite{challenge3}, while glial and oligodendrocytes nucleus are usually round in appearance but the later one have light rings as, astrocytes have oval shape, endothelial cells are usually slender in structure \cite{challenge2}, while malignant tumor cells have irregularly shaped nuclei \cite{challenge4}. Developed model is supposed to be robust enough to detect all these kinds of nuclei without any mislabelling.   Digital image quality variation, background clutter, image artifacts are some other extremely important yet least discussed problem in this domain. Nuclei curvature variation also impacts detection since normally in pathology images, nuclei contour points curvature changes smoothly \cite{challenge5}. However, given a single contour having two or more touching or  occluding nuclei, results large curvature change at touching points. Already existing automated medical image analysis tool use classical segmentation including active contour models, watershed or thresholding  techniques for nuclei instances identification. These tools needs configuration with respect to each data to accurately analyze distinct microscopic modalities including scales and experimental variation, thus leading to an evident requirement of technological domain expertise for accurate algorithm selection and parameters adjustment. Still for proficient ones too, this choice can be daunting,  considering that every year numerous papers are published, presenting new research techniques for gland and nucleus instance segmentation. Even after examining under controlled experimental conditions, no single technique can be generalized for segmenting  all microscopy images correctly, since classical machine learning algorithms are either sensitive to technical artefacts or often fails in adapting to biological samples  heterogeneity. Altogether, this situation yields slows pace of research and at times  inhibits research laboratories from adopting newer image analysis technologies owing to the time and expertise required. 

Many methods have been proposed for the task of nuclei and glands instance segmentation and broadly these can be divided into two major categories: handcrafted feature extraction approaches and deep learning based approaches. Instance segmentation techniques can be divided into major two approaches including image understanding approach based on pixels as well as artificial features utilization including shape, size, color, and texture for  fulfilling instance segmentation. This majorly targets heuristic methods including K-means clustering,  adaptive thresholding i.e local thresholding method for unevenly illuminated images, active contour model, watershed, graph-cuts and other morphological operations  while the second approach is learning based approach that uses different variants of deep learning based techniques primarily based on neural networks.


\textbf{\subsection{ \textbf{Motivation of the Review }}}

Cell nuclei segmentation in whole slide images is often considered as the stepping stone towards whole slide image analysis in any biomedical and biological applications.
The quality of chronic disease diagnosis, survival prediction, phenotype classification, feature extraction and cell tracking highly depends on segmentation accuracy of instances. It gives us the challenging opportunity to study tissue and cellular phenotyping and to draw biological conclusions at large scale.
In recent years, challenges organized specifically for nuclei and gland instance segmentation has definitely brought significant improvement, including public accessibility of large annotated segmentation dataset and 2-D architectures extension for 3-D nuclei images. As compared to natural objects, nuclei detection and instance segmentation seems  easier and simpler due to homogeneous properties in representation. However, despite research on nuclei detection and segmentation topic for decades, still there is no publicly available model supporting nuclei instance segmentation and detection across whole slide images (WSIs) from various labs generated at different conditions. As described by Xu et al. \cite{8759530} till date, even today, this debate towards a generalized architecture or a bench mark solution for all types of  nuclei image segmentation is alive. Before  arrival of convolutional neural networks, conventional nuclei segmentation methods were based on either geographical or statistical image features for seeds generation. A major domain shift from hand crafted feature segmentation technique to using CNN variants for feature extraction has also been observed. This review has been conducted with an aim to provide a comprehensive overview of techniques being used in last 5 years for nuclei and glands instance segmentation tasks as well as identification of most used approach amongst all as a common interactive method for instance segmentation.

\subsection{\textbf{Scope of the Review}}

For this survey we have targeted 126 research papers published in top conferences and journals for the span of last 5 years ranging from 2017-2022. For evaluating advantages and draw backs of each segmented technique a critical review is compiled having existing deep learning computational approaches for nucleus instance segmentation, focusing both hand-crafted morphological feature-based techniques as well as deep neural deep neural network-based methodologies. We provide a comprehensive coverage of the major publicly available datasets being used for the task of nuclei and glands instance segmentation, an extensive summary of grand challenges held globally for the task of instance segmentation as well as classification including Data Science BOWL Challenge (DSB-2018), MoNuSeg, MoNuSAC, GlaS, and CoNIC. There are review articles available in kindred domain \cite{angel2017survey, li2022comprehensive}, but this is the first review paper that comprehensively covers most recent approaches being developed for the task of both nuclei and glands instance segmentation. Through this critical analysis we aim to provide a recapitulation of nuclei and gland instance segmentation techniques for fatal disease diagnosis via integrating automated tools and complex semantic networks. We have discussed some of the existing challenges being faced during analysis, including varying staining impacts, insufficient data causing over fitting, disparate nuclei and glands structure, models specificity to a single image set. Also, we have highlighted key challenges as well as major problems along with outlining future directions. Finally, we present potential future possibilities in generalizing this pivotal task.

\subsection{\textbf{Comparison with other Reviews}}
\begin{figure}[H]
\centering
\includegraphics[ width=1 \textwidth, height = 8cm]{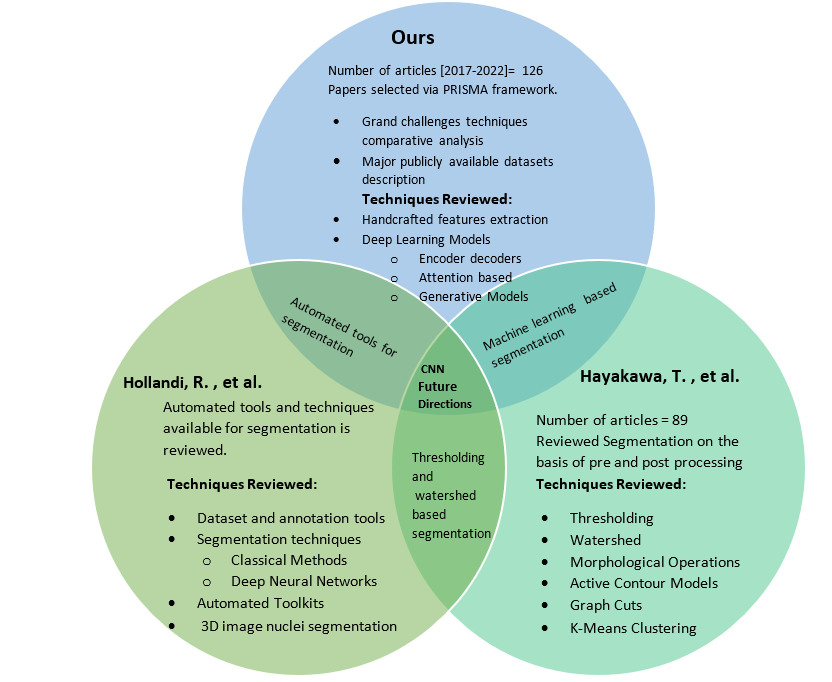}
\caption{Comparison with Recent Reviews}\label{fig3}
\end{figure}

A survey published in 2019 by Hayakawa et al. \cite{review} on computational nuclei segmentation methods in digital pathology discusses major challenges faced in digital pathology and nuclei segmentation due to staining variations during slides preparation  and image modality as well as morphological differences in nuclei. Majors categorization was on the basis of pre and post processing and techniques discussed seeds detection, color normalization, thresholding, watershed, active contour models, graph cut and kmeans. It is  similar to our study  in terms of key element (nuclei instances) and partially for techniques (deep learning and machine learning).
However, 82 papers were reviewed in their study while we review 126 papers.
Another recent survey in 2022 by Hollandi et al. \cite{hollandi2022nucleus} for  nucleus segmentation. They have provided an overview of currently available datasets and annotation tools for training and testing models.Further pre and post processing techniques and challenges related to each is briefly described followed by insights into nuclei segmentation available automated tools and methods covering both classical approaches as well as deep neural based models. For catering issues faced in 3D WSI processing a set of most successful methods yielding promising results are discussed as well.
 We have compared main points in Figure \ref{fig3}.

This review related to our study in terms of context (nuclei segmentation) however, it was  majorly focusing on automated tool kits available for segmentation with a brief summary of methods used while we will be focusing on generalization of techniques adopted in the span of these 5 years with respect to challenges held and datasets used. For comparative analysis of our review with Hayakawa et al. \cite{review} and Hollandi et al. \cite{hollandi2022nucleus}.

\section{Survey Methodology}
In this section, we have discussed the study selection criteria for selection of articles reviewed in this paper. We have also segregated articles according to publication databases (journals or conferences), types (nuclei, gland), impact factor and technique for segmentation (handcrafted feature extraction, deep neural network learning). We have segregated learning-based techniques into the following categories on the basis of their technical contributions:
\begin{enumerate}
\item  Hand crafted segmentation techniques
\item Deep learning based segmentation
\begin{itemize}
 \item Encoder decoder based segmentation
 \item Region based segmentation
 \item Adversarial models based segmentation
 \item Attention-based segmentation
 
 \end{itemize}
   
\end{enumerate}

\subsection{\textbf{Study Selection }}
\begin{figure}[H]%
\centering
 \adjustbox{trim=0 12cm 0 0}{
      \includegraphics[width=1\textwidth]{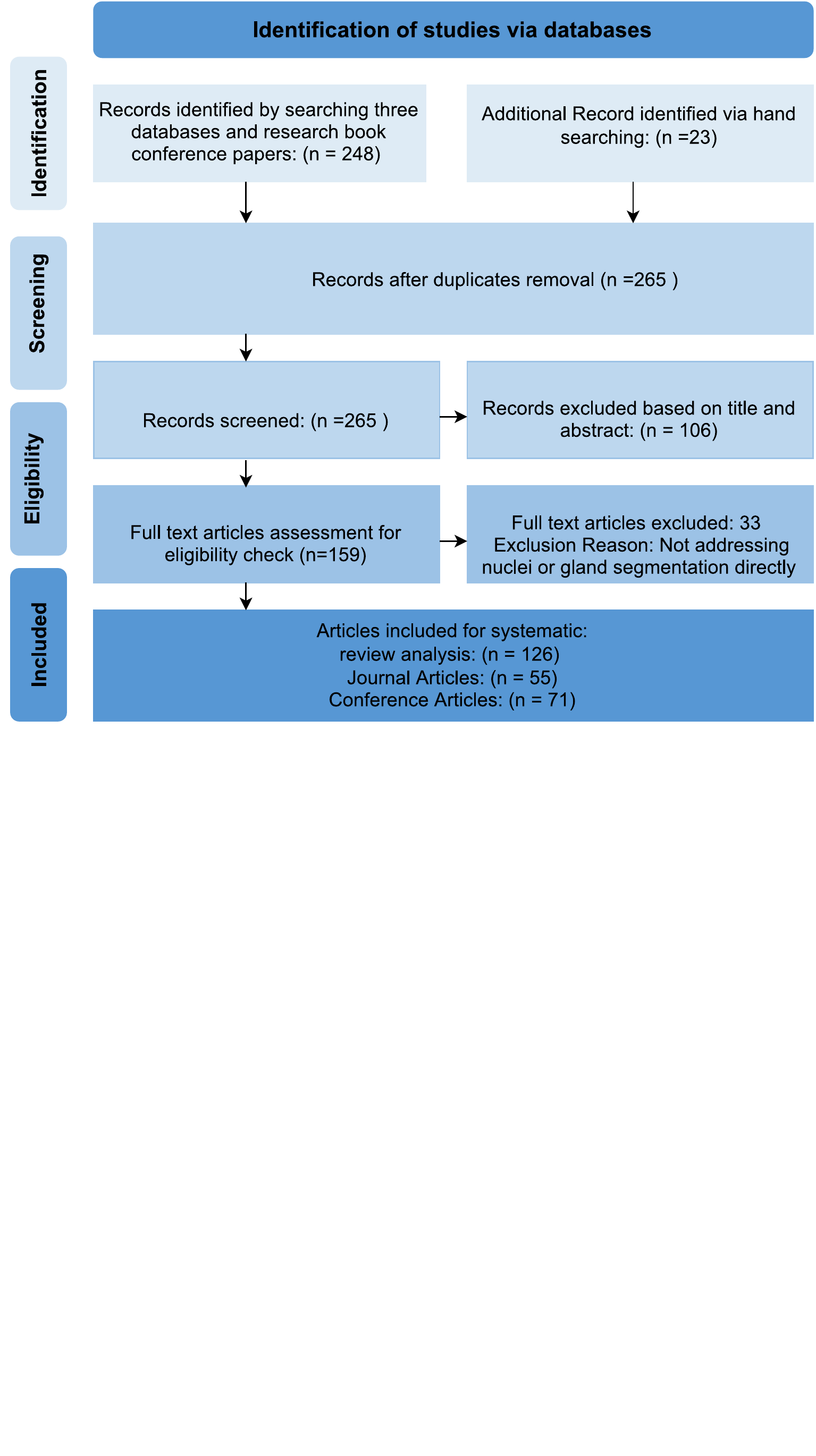}}
        
\caption{PRISMA- Papers Selection Process Summary}\label{fig4}
\end{figure}

We have mentioned quantitative measures of articles searched  and included for review  through preferred reporting items for systematic review and meta analysis (PRISMA) criteria. Figure \ref{fig4} shows papers selection summary. 
We have searched research papers on Springer, Science Direct, IEEE Xplore and lastly Google Scholar. From applying different search queries we have found 261 articles, amongst these 9 were duplicates and are removed. Out of remaining 256 articles, 106 were excluded on the basis of title or abstract. Since they  were not fulfilling the criteria of this review, in some of those articles segmentation of cell was discussed instead of nuclei or gland, while rest of the papers were based on general histology image segmentation. After final full text screening further 46 articles were eliminated for not addressing nuclei or glands segmentation directly. In the PRISMA we have represented papers selection distribution for final 126 included papers. Our PICO question  for this study was: Comparison of all new techniques devised during the last 5 years for nuclei instance segmentation and their efficiency on evaluation for various datasets.

\subsection{\textbf{Data Extraction Methods}}
We targeted top journals and conferences for extracting from different platform including Springer, Science Direct and IEEE Explore for extracting research articles. We extracted deep learning, machine learning and handcrafted based features extraction approaches on nuclei and glands instance segmentation articles published between 2017 and 2022. This specific time frame is chosen due to rapid advancement in techniques proposed for segmentation during this specific tenure and domain shift from classical towards deep networks learning. Also, large number of grand challenges were held from 2017-2022 on histopathological image analysis. We used oriented search string by combining different key words with the logical operators ‘AND’, ‘OR’ to get the relevant papers. Following are the search terms being used for research papers selection.


\begin{itemize}
    \item Nuclei segmentation, Nuclei detection, Gland segmentation, Nuclei instance segmentation, Gland instance segmentation
    \item H\&E,  Whole slide imaging, Pathology, Histopathology
\end{itemize}   
  The inclusion and exclusion criteria for papers is represented in table \ref{inclusion-exclusion}. Initially, papers were selected based on their titles, further abstract, conclusion and methodology is considered for selection where titles did not match inclusion exclusion criteria.


\begin{small}
\begin{longtable}{ |m{6.3cm}| m{6.3cm} |}
 
  \caption{Inclusion Exclusion Criteria.}  \label{inclusion-exclusion} \\
   
 \hline 
        \rowcolor[HTML]{fbebd3}
        \textbf{Inclusion Criteria} &
        \textbf{Exclusion Criteria}  \\
        \hline
\endfirsthead
\endhead
\hline 
\endfoot
\hline 
\endlastfoot

 \rowcolor[HTML]{fdf3e7}
        {Articles published during 2017-2022}& 
        {Articles that were not targeting nuclei instance segmentation directly}\\
        \hline
        \rowcolor[HTML]{fdf3e7}
        {Paper describing detailed techniques and parameters} &
        {Articles based on general medical image segmentation}\\
        \hline
            \rowcolor[HTML]{fdf3e7}
        {Articles having nuclei segmentation, nuclei instance and gland segmentation keywords in title} &
        {Articles that are merely book chapters and not part of any conference journal}\\
        \hline
        \rowcolor[HTML]{fdf3e7}
        {Papers describing datasets} &
        {Papers published in conferences other than mentioned one}  \\
  \end{longtable}
 \end{small}

\begin{figure}[H]%
\centering
\includegraphics[width=1\textwidth, height=8cm]{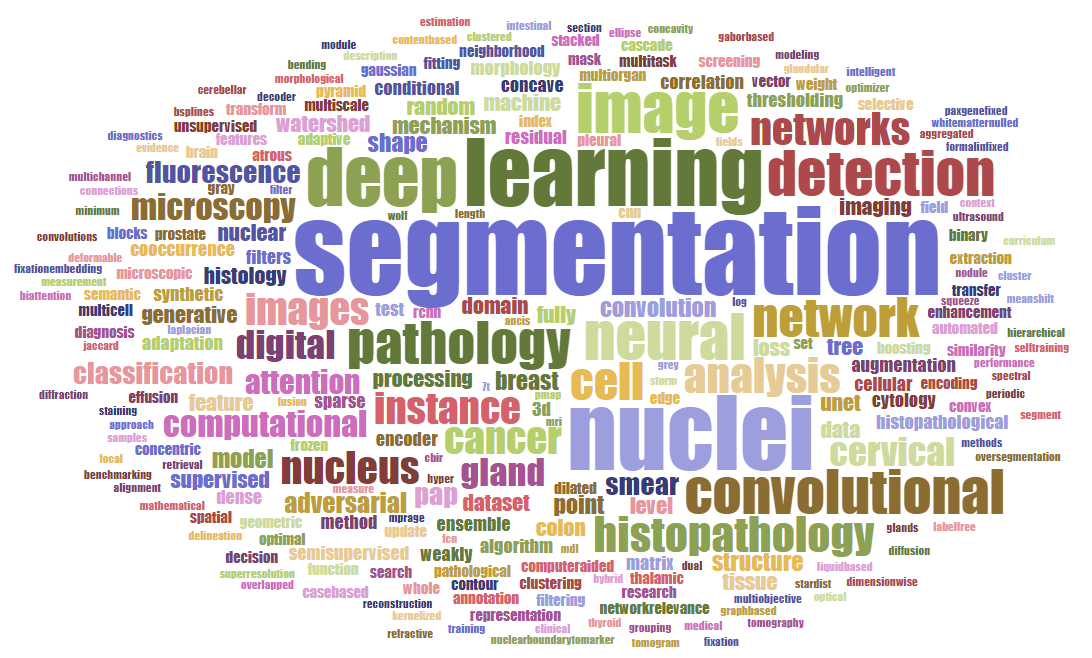}
\caption{A visual depiction of most frequently used keywords in the reviewed articles  }\label{fig5}
\end{figure}

 We collected the keywords from all the reviewed article included in our review and generated the word cloud which is shown is Figure \ref{fig5}.
 
 As our primary objective in this review is to summarize the techniques used for nuclei and glands segmentation, this tag cloud majorly highlights (pathology, histology, deep learning, nucleus, gland, segmentation, cancer detection, classification) terms.


\subsection{\textbf{Papers Distribution}}
In this section we have represented the distribution of papers across various journals, conferences, impact factors, and nuclei or glands types. Major of these division statistics is to give an eagle eye view of published research work, during last 5 years and amount of variation.

\begin{figure}[H]
\centering
\includegraphics[width=0.95\textwidth, height=7cm ]{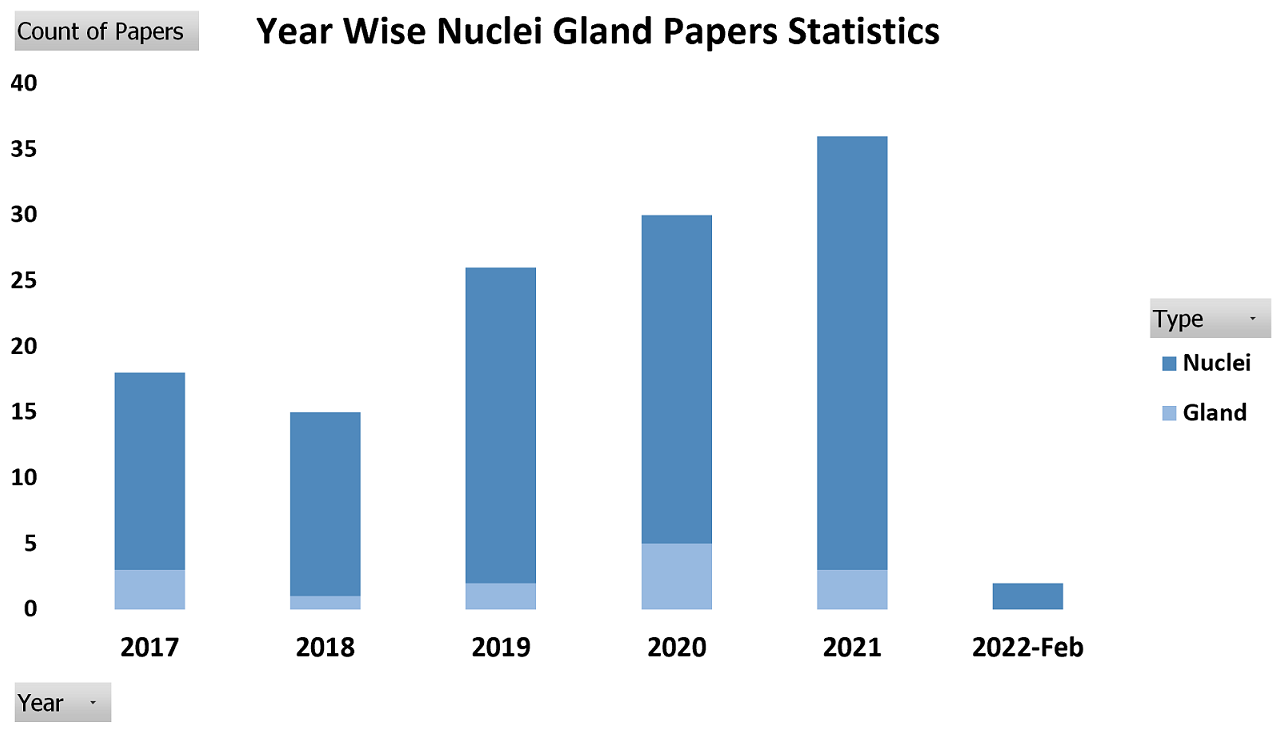}
\caption{Stacked bar chart indicating year wise papers statistics for nuclei and gland }\label{fig6}
\end{figure}

The figure \ref{fig6} shows reviewed articles distribution in last few years. It can be analyzed from the bar chart that the literature for nuclei and glands instance segmentation  has seen a spike throughout the years from 2017 to 2022 since the rise of medical imaging competitions for better diagnostic and prognostic techniques.

\begin{figure}[H]
\centering
\includegraphics[width=0.95\textwidth]{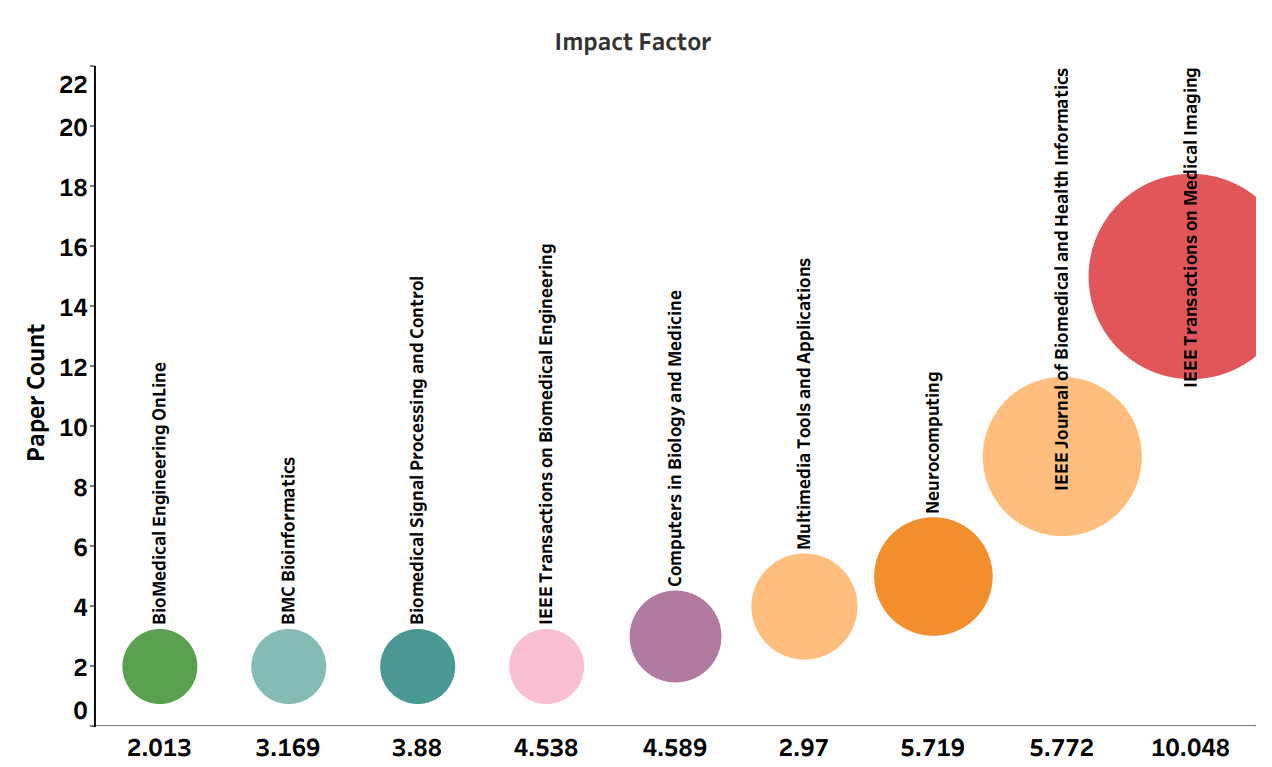}
\caption{ Visual representation of selected research publications from top tier journals along with their impact factor}
\label{fig7}
\end{figure}

 Figure \ref{fig7} shows Impact Factor wise papers distribution. Vertical axis shows the count of papers for each Impact Factor while horizontal axis shows impact factors.  

The distribution of papers per source is shown below in Figure \ref{fig8}. From the figure count of nuclei and glands based medical image segmentation articles taken from various top journals and conferences can be depicted. Each bubble represents the number of papers taken from a specific journal or a specific conference.
\begin{figure}[H]
\centering
\includegraphics[width=8cm]{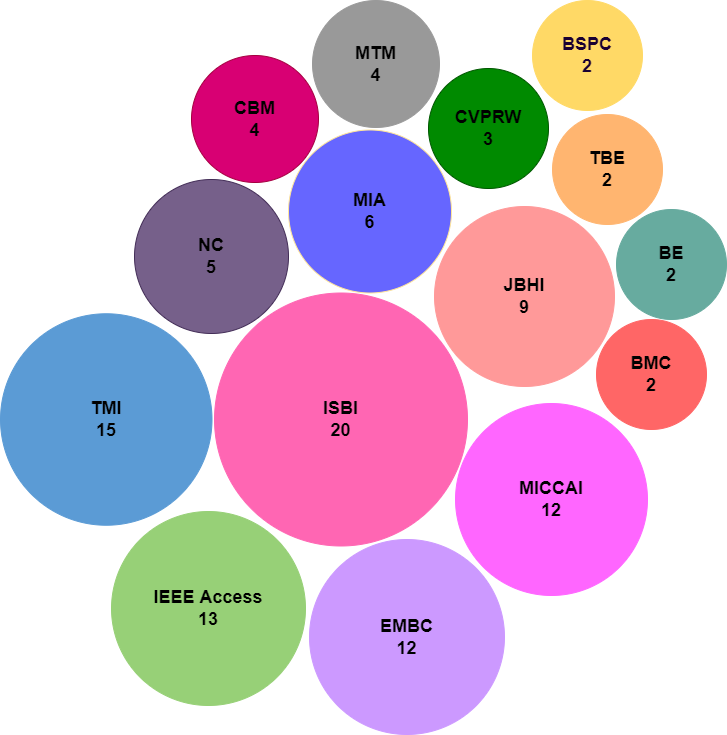}
\caption{Bubble graph representing number of articles chosen from top ranked journals and conferences }\label{fig8}
\end{figure}

\section{ Datasets}

The dataset reviewed in this study covers a wide range of image sets featuring prominent open-source image databases from the PanNuke, Multi-Organ Nucleus Segmentation and Classification (MoNuSAC), CryoNuSeg, Lizard to small private datasets thus enhancing diversity of research study. Major tissues included in datasets includes breast, liver, bladder, colon, stomach, lung, kidney, prostate, cervix, gland and brain. Table \ref{table2} summarize all the details of the given datasets including their source.

\begin{figure}
    \centering
    \begin{tabular}{cc}
        \subfloat[]{\includegraphics[width=3.1cm,height=3.1cm]{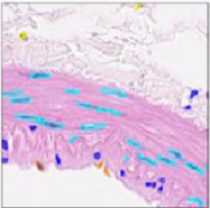} 
        \includegraphics[width=3.1cm,height=3.1cm]{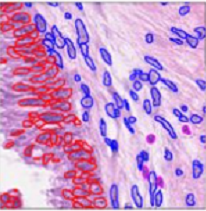}} &
        
        \subfloat[]{\includegraphics[width=3.1cm,height=3.1cm]{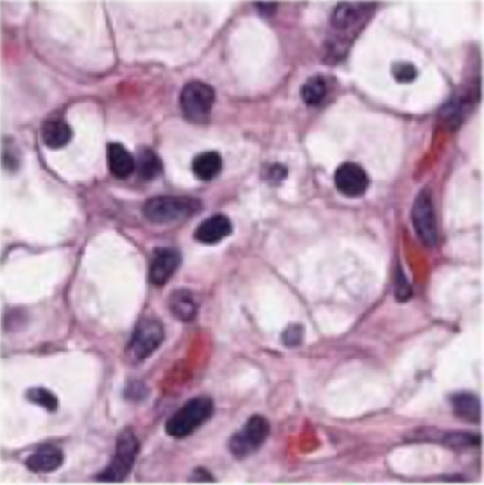} 
        \includegraphics[width=3.1cm,height=3.1cm]{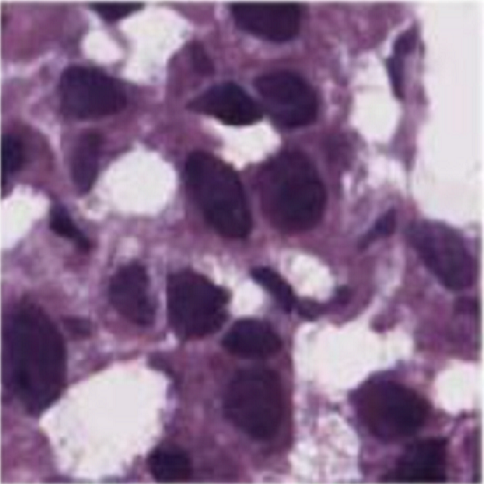}} \\
        
        \subfloat[]{\includegraphics[width=3.1cm,height=3.1cm]{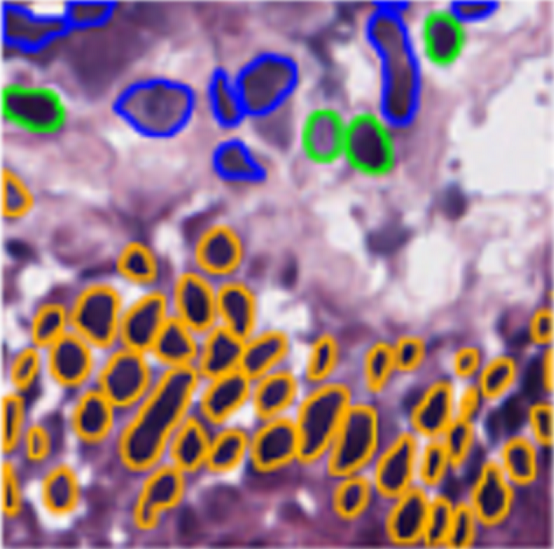} 
        \includegraphics[width=3.1cm,height=3.1cm]{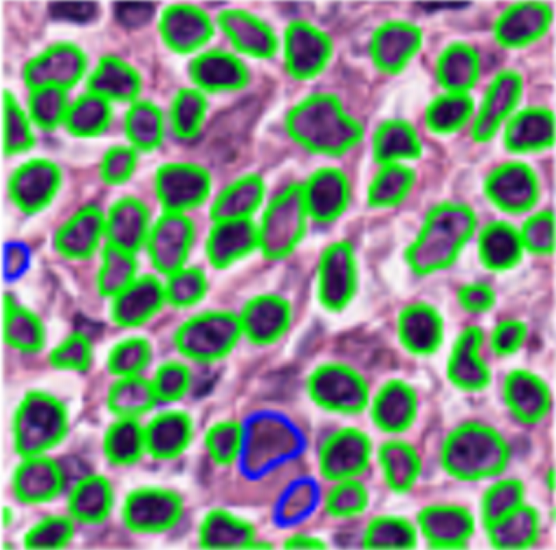}} &
        
        \subfloat[]{\includegraphics[width=3.1cm,height=3.1cm]{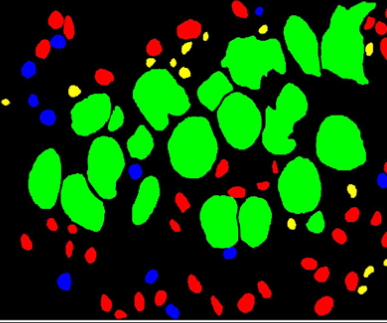} 
        \includegraphics[width=3.1cm,height=3.1cm]{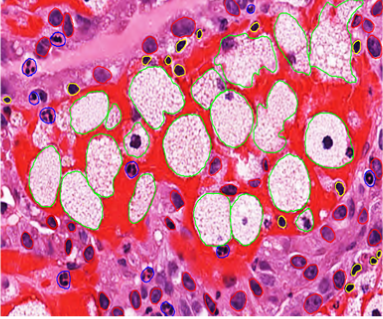}} \\
        
        \subfloat[]{\includegraphics[width=3.1cm,height=3.1cm]{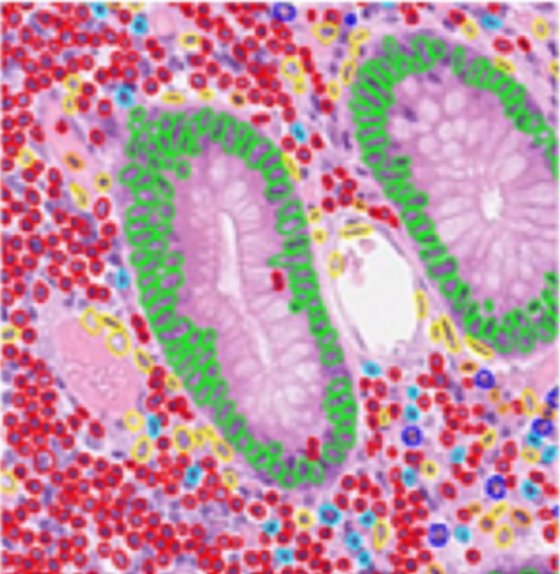} 
        \includegraphics[width=3.1cm,height=3.1cm]{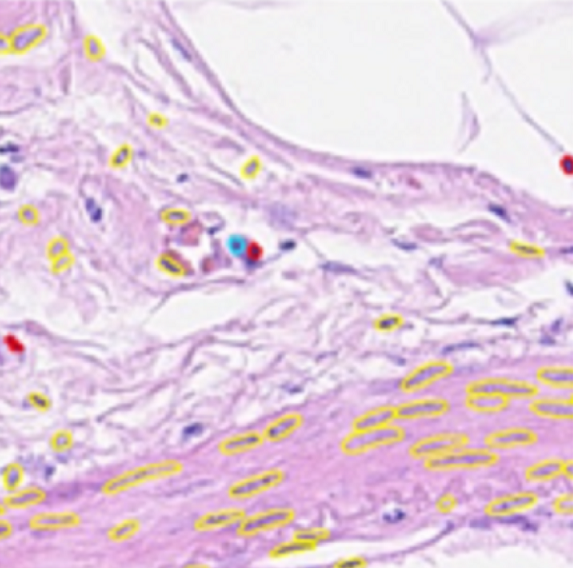}} &
        
        \subfloat[]{\includegraphics[width=3.1cm,height=3.1cm]{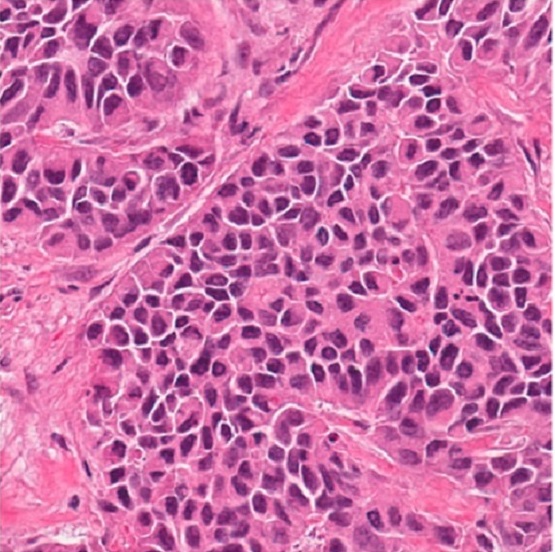} 
        \includegraphics[width=3.1cm,height=3.1cm]{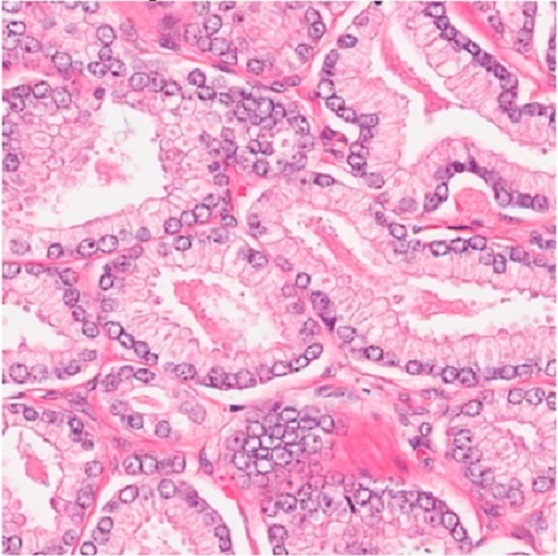}} \\
        
        \subfloat[]{\includegraphics[width=3.1cm,height=3.1cm]{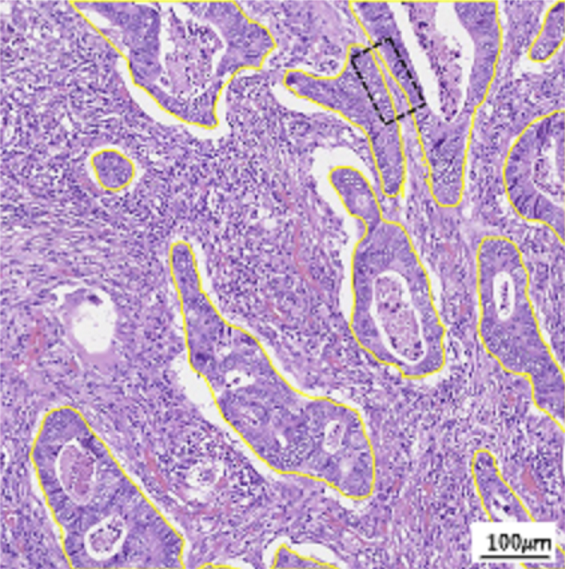} 
        \includegraphics[width=3.1cm,height=3.1cm]{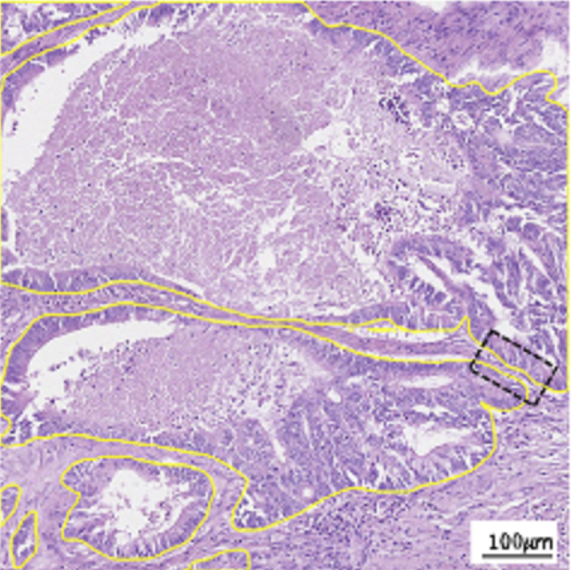}} &
        \subfloat[]{\includegraphics[width=3.1cm,height=3.1cm]{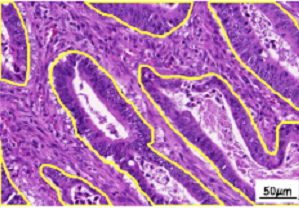} 
        \includegraphics[width=3.1cm,height=3.1cm]{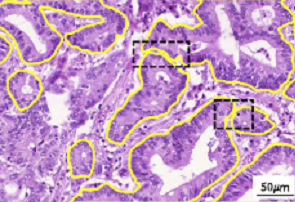}} \\
    \end{tabular}
    \caption{A catalogue of morphological variation in  nuclei at 20x magnification for different organs.  \textbf{(a)} CoNSeP \textbf{(b)} Kumar \textbf{(c)} PanNuke \textbf{(d)} MoNuSAC \textbf{(e)} Lizard \textbf{(f)} CryoNuSeg \textbf{(g)} CRAG  \textbf{(h)} GlaS}
    \label{datasets}
\end{figure}

\subsection{\textbf{Nuclei Datasets}}

\subsubsection{\textbf{CoNSeP}}
CoNSeP (Colorectal Nuclear Segmentation and Phenotypes Dataset) is one of the first fully annotated  open source dataset that enabled the development of models for simultaneous nuclear segmentation and classification approaches in Computational Pathology. It consists of 41 Haematoxylin and Eosin (H\&E) images of size 1,000×1,000 px from colorectal adenocarcinoma (CRA) Whole slide images (WSIs) \cite{yen2020ninepins}. 

\subsubsection{\textbf{Kumar}}
Kumar dataset contains 30 H\&E stained image tiles of 1,000×1,000 px size from seven different organs including breast, liver, kidney, prostate, bladder, colon and stomach  acquired from The Cancer Genome Atlas (TCGA) database at 40× magnification. Boundaries of each nuclei within that image  are fully annotated. 

\subsubsection{\textbf{PanNuke}}
PanNuke \cite{inbook} (An Open Pan Cancer Histology Dataset for Nuclei Instance Segmentation and Classification) is the biggest and the most diverse semi-automated datasets created till date for nucleus instance segmentation task. Its nuclei boundaries have been created automatically and further validation is performed by experienced pathologists. This is formulated by formalin fixed and paraffin embedded samples having about 200,000 nuclei spanning over 19 different organs comprising of 481 visual fields forming patches of size 256 x 256 at different magnifications. These patches are further randomly distributed into training, testing and validation subsets. It is of significant clinical importance because of its clinical importance specifically with respect to tissue phenotyping and technical significance as well.

\subsubsection{\textbf{MoNuSAC}}
The multi organ nucleus segmentation and classification (MoNuSAC) \cite{9446924} is a large and diverse dataset having boundary annotations for each nuclei along with class labels. Previously released public datasets for nucleus segmentation were either not having multiple organs data or are not curated to the level of individual nuclei. It comprises of  over 46,000 hand-annotated nuclei from 71 patients spanning over 31 hospital and four organ cell types including epithelial, macrophages, lymphocytes and neutrophils.


\subsubsection{\textbf{Lizard}}
Lizard \cite{lizard} (A Large-Scale Dataset for Colonic Nuclear Instance Segmentation and Classification) is the  largest known dataset available for nuclei segmentation and classification in digital pathology having nearly half a million annotated nuclei along with their class label, consists of whole slide image regions of colon tissue having six different dataset sources including  GlaS, CRAG, CoNSeP, Digest Path, PanNuke and TCGA at 20× magnification. Its fully annotated dataset for about 495,179 nuclei. In particular we provide the nuclear class label for epithelial cells, connective tissue cells, lymphocytes, plasma cells, neutrophils and eosinophils.

\subsubsection{\textbf{CryoNuSeg}}

CryoNuSeg \cite{2021} is the first fully annotated  H\&E stained dataset formulated by frozen samples (FS) derived images and the organs images used in dataset compilation have not been a part of any prior dataset. It comprises of 30 digitized H\&E stained images derived from 10 different organs. Tissue types includes gland, larynx, adrenal, lymph nodes, pancreas, skin, pleura, mediastinum, thyroid gland, thymus, testes.
It can be used as a stand alone benchmark dataset or in combination with other publicly available datasets for training supervised machine learning and deep learning models.

\subsection{ \textbf{Glands Datasets}}

\subsubsection{\textbf{GlaS}}
Gland Segmentation Challenge dataset (GlaS) is first used as part of  MICCAI-2015. This data is extracted from 16 Haematoxylin and Eosin H\&E stained Whole slide images (WSIs) scanned through MIRAX MIDI Slide Scanner pixel resolution at 20 x magnification.
It consists of total 165 images out of which 85 are used as training (48 malignant and 37 benign) and 80 test images including (43 malignant and 37 benign). Size of images is 775 x 522 pixels each having associated instance-segmented ground truths highlighting glands boundaries align with accurate lumen annotations for glands.


\subsubsection{\textbf{CRAG}}
CRAG (Colorectal Adenocarcinoma Glands) dataset is primarily comprised of gland images. It consists of colon adenocarcinomas usually referred as colorectal gland (CRAG) dataset  developed by University Hospital Coventry and Warwickshire (UHCW) NHS Trust Coventry UK. It composed of 213 H\&E stained CRA images from 38 whole slide images (WSIs) scanned by VL120 scanner at 20 x objective magnification and are mostly of size 1512 x 1512 px along with instance level ground truth. Training and testing  images are 173 and 40 specifically with varying cancer grades.

\begin{tiny}
\begin{longtable}{|m{0.5cm} | m{1cm} |m{1.1cm} | m{1.9cm}  | m{1cm} | m{1.2cm} | m{2.5cm} | m{1.1cm} |}
 
  \caption{Nuclei and Glands Segmentation Datasets in public domain} \label{table2} \\

\hline
    \rowcolor[HTML]{FBEBD3}
    \textbf{S.No} &
    \textbf{Dataset} &
    \textbf{\# Nuclei}&
    \textbf{Label Type} &
    \textbf{Magnifi- cation} &
    \textbf{\# Organs} &
    \textbf{Organs} &   
    \textbf{Source}
    \\
    \hline
\endfirsthead

\multicolumn{3}{c}%
{{ }} \\
\hline 

\endhead

\hline 

\endfoot

\hline 
\endlastfoot
\rowcolor[HTML]{fdf3e7}
1 & 	CoNSeP &	24319 &	Instance + Classification &	40x &	1 &	Colon &	UHCW$^2$
\\ \hline
       \rowcolor[HTML]{fdf3e7}
2 & 	Kumar &	21623 &	Instance &	40 x &	7 &	Breast, Liver, Kidney, Prostate, Bladder, Colon, Stomach &	TCGA$^1$
\\ \hline
     
        \rowcolor[HTML]{fdf3e7}
 3 & 	MoNuSAC &	46909 &	Instance + Classification & 40x &	4 &	Breast, Kidney, Lung, Prostate &	UHCW$^2$
\\ \hline
       \rowcolor[HTML]{fdf3e7}
4 &  Lizard &	495,179	& Instance + Classification & 20x &	1 &	Colon &	TCGA$^1$
\\     \hline

         \rowcolor[HTML]{fdf3e7}
5 & 	GlaS &	n/a &	Instance &	20x &	2 &	Colon, Prostate	& n/a \\ \hline
                 \rowcolor[HTML]{fdf3e7}
6 &	CRAG & n/a &		Instance &	20x &	1 &	Colon &	UHCW$^2$
\\     \hline

  \rowcolor[HTML]{fdf3e7}
   7 & 	PanNuke &	205,343 &	Instance + Classification &	40x &	19 &	Adrenal, Bile-duct, Bladder, Breast, Colon, Cervix, Esophagus, Head, Neck, Kidney, Liver, Lung, Ovarian, Pancreas, Prostate, Skin, Testis, Stomach, Thyroid, Uterus	& TCGA$^1$
    \\ \hline
    
        \rowcolor[HTML]{fdf3e7}
8 & 	CryoNuSeg &	7596 &	Instance+ Classification &	40x &	10 & Adrenal gland, Larynx, Lymph node, Mediastinum, Pancreas, Pleura, Skin, Testis, Thymus, Thyroid gland & TCGA$^1$
\\     \hline

  \end{longtable}
   \end{tiny}
\footnotetext[1]
{ \href{https://warwick.ac.uk/fac/cross_fac/tia/data/glascontest/abot/}{The Cancer Genome Atlas}}

\footnotetext[2]
{ \href{https://warwick.ac.uk/fac/cross_fac/tia/data/glascontest/about}{University Hospital Coventry \& Warwickshire
}}

\begin{figure}[H]%
\centering
\includegraphics[width=10cm, height=7cm]{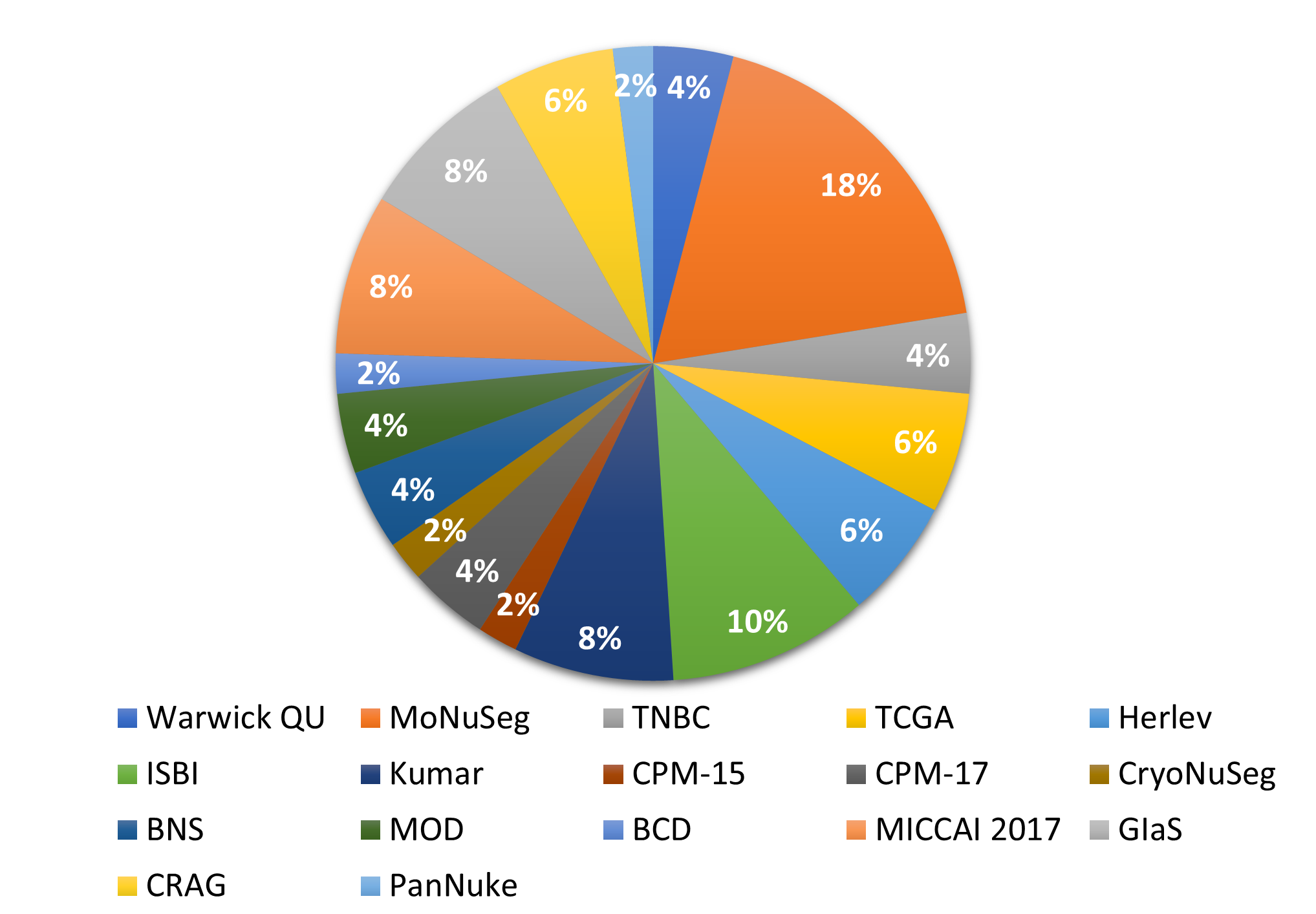}
\caption{ Reviewed articles wise percentage distribution of nuclei and glands datasets }\label{fig10}
\end{figure}

\begin{figure}
\centering
\includegraphics[width=1\textwidth, height=8cm]{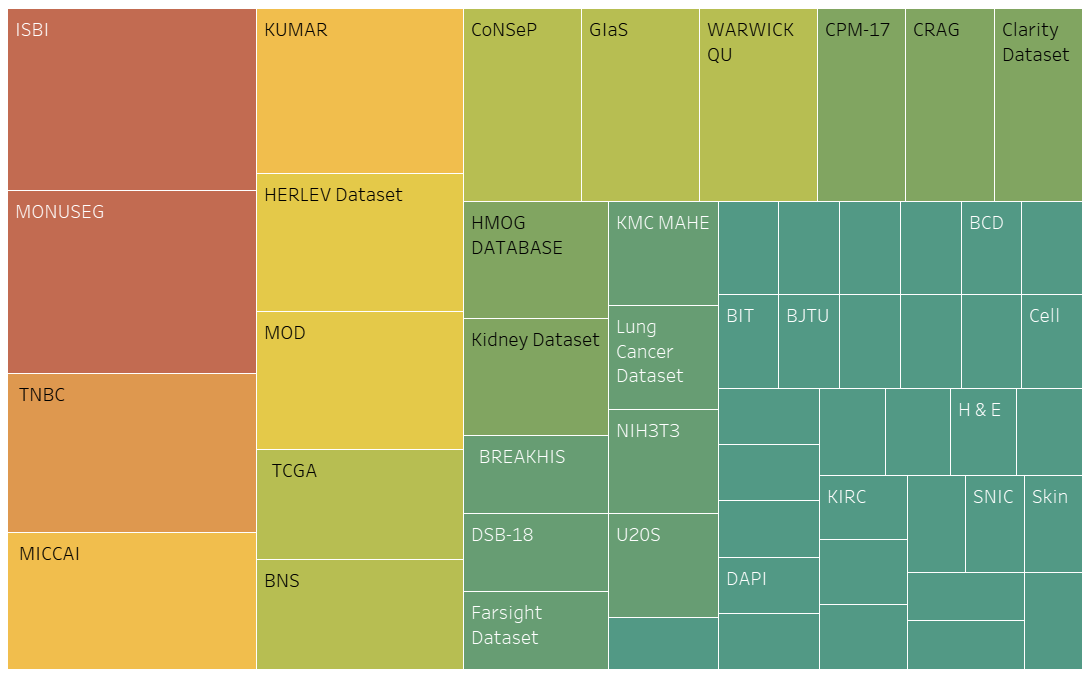}
\caption{ Tree map representing frequency of datasets and its sources }\label{fig9}
\end{figure}

Figure \ref{fig10} shows percentage occurrence of nuclei and glands publicly available datasets used in the reviewed articles. The graph shows that most of the nuclei instance segmentation papers have used MoNuSeg, ISBI, MICCAI challenge and Kumar covering 18\%, 10\%, 8\% and 8\% respectively. Similarly, glands instance segmentation articles have used Warwick-QU, CRAG and GlaS datasets comprising over 8\%, 8\% and 6\% distribution respectively. Similarly, figure \ref{fig9} shows datasets usage frequency from various sources in the reviewed articles.

\begin{figure}[H]
\centering
\includegraphics[width=1\textwidth, height=10cm]{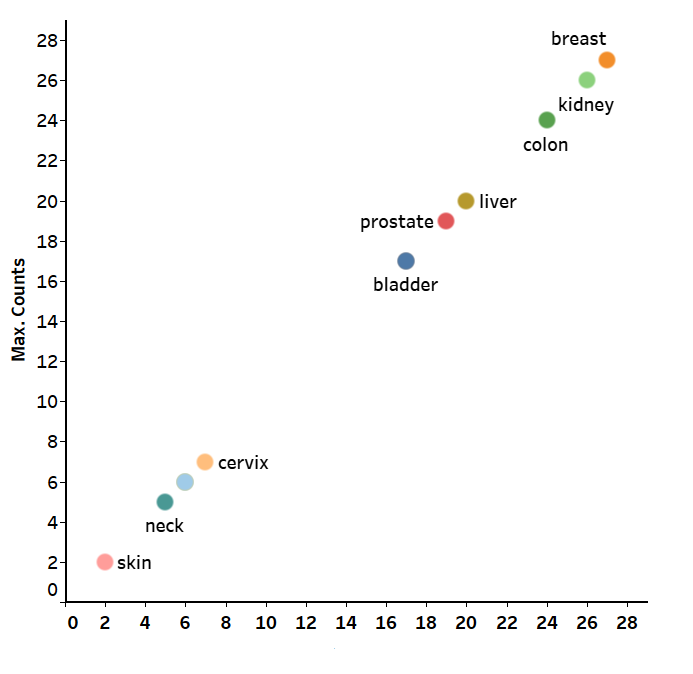}
\caption{ Most frequently used tissue slides in research articles }\label{fig11}
\end{figure}

Figure \ref{fig11} depicts most frequently used tissue slides for the task of nuclei and glands segmentation within publicly available datasets or via privately generated whole slide images. The plot shows that most of the research has been done on breast, kidney and colon datasets with a count of 28, 27 and 25 simultaneously out of total reviewed. Similarly, prostate, liver and bladder have also been utilized for research purposes. Finally, cervix, neck and skin are the least used ones with occurrence in 8, 6 and 3 papers. 

\section{Grand Challenges}\label{sec3}
\begin{figure}[H]
\centering
\includegraphics[width=1\textwidth]{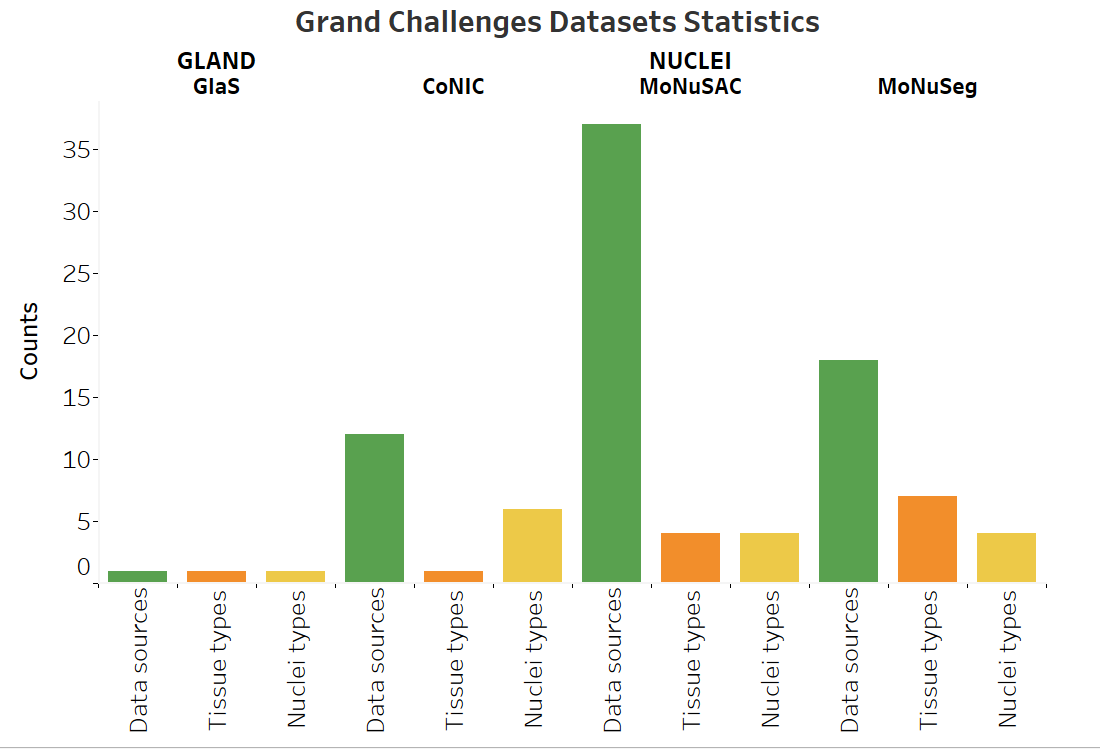}
\caption{The bar chart illustrates precise summary of data collection, tissue and nuclei counts specific to dataset launched for above mentioned grand challenges. Here data  sources counts of various distinct hospitals, universities and databases referred for data collection. Similarly, Tissue type shows distinct variety of tissues in each dataset i.e. kidney, breast, colon. Lastly, Nuclei type count is depicting variation of nuclei kinds specific to each dataset.   }\label{fig12}
\end{figure}

 Grand Challenges have always been an effective approach towards crowd sourcing the development of best performant algorithms as well as pointing out new research directions. Despite a lot of effectiveness towards techniques enhancement and in facilitating innovation, these competitions also suffers from a set of limitations. Most of the times validations of the resulting algorithms have not been typically performed independently by the algorithm developers which later on leads towards a technically bias techniques not reproducible for a general setup. This lack of algorithms validation also poses great doubt towards generalization capability to cater underlying critical issue, rather than merely fine tuning for a particular competition design setups. Figure \ref{fig12} shows data collection statistics from various data sources, number of organs and types used in each grand challenge. Table \ref{table3} shows the dataset details that are released for grand challenges.

\subsection{ \textbf{Nuclei Segmentation Challenges}}
\subsubsection{\textbf{MoNuSeg}}
Multi Organ Nucleus Segmentation Challenge was organized for reducing time to develop and validate visual bio markers for new whole slide image datasets. Preprocessing techniques used by participants includes color normalization being used by most of the participants while Unit Variance, Range Standardization and Histogram Equalization have also been used by some of the participants. Similarly, Segmentation techniques used includes U-Net\cite {ronneberger2015u}, Mask-RCNN \cite{he2017mask}, FCN \cite{long2015fully}, FPN \cite{lin2017feature}, PANet, ResNet \cite{he2016deep}, VGG-Net \cite{https://doi.org/10.48550/arxiv.1409.1556}, DenseNet \cite{iandola2014densenet} and Distance Map. Cross Entropy and Dice Loss are the most used loss functions.

\subsubsection{\textbf{MoNuSAC}}
Multi-Organ Nuclei Segmentation and Classification Challenge has been organized with an aim of detecting, segmenting, and classifying different types of nuclei around the tumor matrix  and it holds a special importance in characterizing the tumor micro-environment for cancer research and prognostication thus freeing up pathologists time for other major tasks. This also helps in reducing chances of errors being caused while doing this task manually. For this challenge a dataset having 46000 nuclei of 4 different organ and types have been used. Majority of participant have used following techniques for achieving better results including pre-processing by color normalization, data augmentation, major CNN architectures used includes U-Net \cite{ronneberger2015u}, FPN \cite{lin2017feature}, FCN \cite{long2015fully}, HoVer-Net \cite{graham2019hover}, DenseNet \cite{iandola2014densenet}, VGG-Net \cite{https://doi.org/10.48550/arxiv.1409.1556} and Efficient-Net \cite{tan2019efficientnet} followed by watershed, morphological operations and thresholding techniques for post processing to fatigue and subjectivity.

\subsubsection{\textbf{Data Science Bowl 2018}}
Around 3890 teams all over the globe participated in this first ever challenge targeting nuclei segmentation. Top participants succeeded in developing a deep algorithm that can be applied to any 2-D image \cite{caicedo2019nucleus}. For this competition around 37,333 manually annotated nuclei from a set of 841 images have been generated after 30 different experiments for varying samples. Majority of participants have used deep convolutional neural networks (DNN's) based techniques due to its better results for  various microscopy image as well as pathology problems. Different variants of CNN architectures have been designed for accurate image segmentation and improving accuracy. Techniques used  by majority of participants include ensemble of U-Net \cite{ronneberger2015u}, fully
convolutional neural network \cite{long2015fully}, Mask-RCNN (Region-based CNN) model \cite{he2017mask} and feature pyramid network (FPN) \cite{lin2017feature}.

\subsection{\textbf{Gland Segmentation Challenge}}

\subsubsection{\textbf{GlaS}}
Gland Segmentation in Colon Histology Images (GlaS) \cite{glascontest} challenge has been organized by MICCAI-2015. Major objective of this competition was solving gland segmentation problem in Haematoxylin and Eosin (H\&E) stained images. Algorithms developed by participant were applied to both  colonic carcinomas and benign tissues. Around 165 images derived from 16 H\&E stained histological slides of stage T3 or T42 colorectal adenocarcinoma have been used for this competition. Best performing algorithms in this competitions have used Fully Convolution Neural Network(FCN) \cite{long2015fully}, Contour based U-Net \cite{ronneberger2015u}, Object-NET, MSER based techniques as well as data augmentation and pre-trained models have been utilized as well. 

\begin{tiny}
\begin{longtable}{|m{0.5cm} | m{2.2cm} |m{1.3cm} | m{1.5cm}  | m{1.5cm} | m{0.8cm}  | m{1cm} | m{1.7cm} |}
 
  \caption{Nuclei and Gland Competition Statistics} \label{table3} \\

\hline
        \rowcolor[HTML]{fbebd3}
        \textbf{S.No} &
        \textbf{Reference} &
        \textbf{Year} &
        \textbf{Challenge}&
        \textbf{Participants} &
        \textbf{Organ} &
        \textbf{Type} &
        \textbf{No. of objects} \\
    \hline
\endfirsthead

\multicolumn{3}{c}%
{{ }} \\
\hline 

\endhead

\hline 

\endfoot

\hline 
\endlastfoot
  \rowcolor[HTML]{fdf3e7}
     1 & Verma et al.\cite{9446924} &  2020 & MoNuSAC  & 170   & 4     & Nuclei & 46000 \\ \hline
           \rowcolor[HTML]{fdf3e7}
      2 & Juan et al. \cite{caicedo2019nucleus} & 2018 & DS-Bowl & 17929  & 1     & Nuclei & 37333 \\ \hline
           \rowcolor[HTML]{fdf3e7}
       3 & Kumar et al. \cite{8880654} & 2018 & MoNuSeg  & 80    & 7     & Nuclei & 21623 \\ \hline
            \rowcolor[HTML]{fdf3e7}
       4 & Graham et al. \cite{7012} & 2022 & CoNIC & 520      & 1     & Nuclei & 431913 \\ \hline
           \rowcolor[HTML]{fdf3e7}
       5 & Gorsuk et al. \cite{glascontest} & 2015 & GlaS & 200       & 1     & Gland  & 52 \\     \hline
   
  \end{longtable}
 \footnotetext[1]
 { \href{https://www.cancer.gov/about-nci/organization/ccg/research/structural-genomics/tcga}{The Cancer Genome Atlas}}

\footnotetext[2]
{ \href{https://warwick.ac.uk/fac/cross_fac/tia/data/glascontest/about}{University Hospital Coventry \& Warwickshire
}}

 \end{tiny}

\section{Nuclei \& Glands Instance Segmentation Methodologies in Histology Images }\label{sec5}

Nuclei \& Glands Segmentation instance segmentation task basically gives information about distinguishing class, location, number of objects and contours in an image.
Considering the need and applications of nuclei and gland segmentation many automated algorithms have been designed for this task. These segmentation approaches can be categorized into two major categories including:  conventional segmentation methods and deep-learning based methods.
Traditional segmentation methods primarily targetting handcrafted image features i.e the variation, gradient, distribution and other color features includes following techniques such as thresholding followed by morphological opening or closing operations, active contour models, graph-based techniques, deformable methods, marker controlled watershed segmentation and their other variants alongside multitudes of other pre and post-processing step addition for achieving segmentation results. Techniques based on deep learning from their ability to learn shape and color variations can achieve better accuracies. In last few years, deep neural networks (DNNs) have rapidly dominated the field of image segmentation and classification. Using the recognizing ability of neural networks, fully convolutional neural networks improves the efficiency of semantic segmentation. Widely speaking, deep learning based nuclei segmentation frameworks can be divided into two major categories. Firstly, CNN-based nuclei detection having deformable models. In this approach,deep neural networks is applied for generating probability map of nuclei centroid while for post processing clustering, watershed or active contour based techniques are sued for nuclei boundary segmentation and algorithms like the watershed transform, clustering, and active contour are utilized to post-process the boundary of nuclei. Second majorly used technique is FCN based pixel wise end-end segmentation. It comprises of an encoder decoder based architecture including fully convolutional layers embedded with a refining technique i.e watershed transform and conditional random field (CRF).

\subsection{ \textbf{Nuclei Instance Segmentation}}
Nuclei instance segmentation play major role toward the automatic diagnosis of cancer and medical image processing and analysis. Plenty of research has been done to efficiently segment out nuclei and gland instances. In this section we have reviewed selected articles segregated primarily in two categories i.e. hand crafted and deep learning based feature extraction techniques. Table \ref{table4} enlist the details of the reviewed articles along with highlighting top performing technique for each category.  

\subsubsection{\textbf{Handcrafted features extraction methods}}
 Variability in size, shape, textural and tissue appearance make the detection of cell nuclei exigent, which is vital for the automatic analysis of digital pathology slides. The analysis of pathology slides is crucial in the quantification of the phenotypic information contained in tissue sections. 
An approach is proposed by Brieu et al. \cite{brieu2017learning} that dealt with the variability in size by treating this detection problem as local maxima detection on center probabilistic map, where a nuclear surface area map with a-priori knowledge on the size of the object of interest used for detecting local maxima. This method exhibited good quantitative displayed.  
Xu et al. \cite{xu2016automatic} presented a generalized Laplacian of Gaussian (gLoG) filter based automatic technique for automatic nuclei detection in digital pathology slides. gLoG filters with different scales and orientation are first piled up and then set of response maps is obtained by performing convolution operation on contender image using directional gLoG kernels. Further they used mean shift algorithm to detect and cluster the local maxima of response map based on special closeness. In each group the point with maximum response is selected as the nucleus. Proposed technique is evaluated on two datasets, shown a good performance in nuclei detection. 
Rojas-Moraleda et al. \cite{rojas2017robust} introduced an abstract simplicial homology approach based on the principles of persistent homology to address the problem of cell nuclei segmentation which identify salient region in the image that exhibit pattern of persistence, by dealing with the persistence of disconnected sets. This topological image representation reduced dependency of the segmentation task on variation of color or texture. Images of liver tissue acquired from histological sections is used to determine the efficiency of the approach. The proposed method recognized hepatocyte and non-parenchymal cell nuclei combinely, with an accuracy of 84.6\%, and hepatocyte cell nuclei alone with an accuracy of 86.2\%. 
Gautam et al. \cite{gautam2017unsupervised} proposed a contrast based adaptable versions of mean-shift and SLIC algorithms for the segmentation in pap smear images. This algorithm is then followed by an intensity weighted adaptable thresholding. The proposed model evaluated using Herlev dataset, achieved effective performance on images having inconsistent contrast in comparison with state-of-the art clustering-based method. 
Similar, accelerated acquisition Diffusion filter based kmeans clustering approach is used by  Battistella et al. \cite{Battistella2016RobustTN} for segmentation on thalamic brain anatomy dataset, model reported 84\% accuracy.
 Reljin et al. \cite{reljin2017multifractal} presented an inverse multifractal analysis (IMFA) for segmentation of nuclei in fluorescence in-situ hybridization (FISH) images. At first the matrix of Holder exponent from blue channel of FISH image, along with one-by-one conformity with the RGB image is determined. The proposed semi-automatic method, initially apply predefined hard thresholding to perform segmentation from the matrix of Holder exponents then segmentation is refined by changing the threshold as a result of user evaluation. The evaluation of the IMFA segmentation method carried out over 100 clinical cases, showed that the benefits of the proposed method compared to already reported methods.
Zhang et al. \cite{ zhang2017automated} presented a four stepped method for robustly and efficiently segmenting overlapped nuclei. These steps are contour extraction, concave point detection, contour segment grouping and ellipse fitting. Contour extraction algorithms are used for the estimation of the level of the image blurriness, which determines parameters in following steps for increasing the segmentation accuracy of the blurry nuclei. Different algorithms are proposed for extracting obvious and unobvious concave from the contender points. Grouping rules are proposed for grouping segments of the concave points.
Kostrykin et al. \cite{challenge6}  proposed a second order optimization technique for intensity based segmentation on fluorescence microscopy images. It targeted only below threshold  haematoxylin intensity based areas within nuclei while all higher areas of intensity in background. It resulted in 75\% accurate segmentation results on NIH3T3 and  ISBI-2013 datasets. Hameed et al. \cite{watershednew3} adopted gray-level co-occurrence matrix based cross entropy thresholding approach via taking histogram of input image for minimizing entropy for optimum threshold value assignment followed by post processing segmented objects via watershed transform this enhanced segmentation DSC score to 96\%. However on gray scale images this thresholding based segmentation resulted in poor DSC score which is improved via pre processing.

Similarly, Quachtran et al. \cite{count} also worked on fluorescence images instance segmentation using iterative radial voting techniques, however there results accuracy is slightly less 64\% dice score compared to optimization based approach.

Lee et al. \cite{lee2017nucleus} identified the contender nuclei seeds as extrema in a Laplacian-of-Gaussian space. Similarly, non-nuclei seeds are removed from clusters acquired by ellipse fitting. Local and global thresholding is combined to define region of interest. shape and roughness of shared boundaries connected nuclei is modeled for repeatedly merging and splitting these regions. The model shows a success in splitting boundaries of connected nuclei and recognizing the nucleus region.   
Saha et al. \cite{saha2018segmentation} detect and segment nuclei by merging over-segmented SLIC super pixel regions using a novel image consolidating technique based on pairwise special contrast and image gradient contour evaluation, from cervical cytology images in ISBI-2014 dataset. First overlapping cervical cytology image segmentation is used for the evaluation of the proposed framework. The framework surpasses the performance from the state-of-the-art detection and segmentation algorithms.
Guo et al. \cite{guo2018clumped} proposed algorithm for segmentation of overlapped nuclei. The algorithm identifies contenders with point pair connection and evaluates abutted point connections with a contrive ellipse fitting quality criterion. After the establishment of the connection relationship, the ensue dividing paths are recovered by following the path of certain eigenvalues from the image hessian in a contrived searching space. Qualitative and quantitative evaluation, carried on 560 image patches from two classes shows the promising results of the algorithm.
Semedo et al. \cite{semedo2018thalamic} proposed an algorithm for the segmentation of the thalamic nuclei, a central part of the nerve propagating the impulses between sub cortical regions and the cerebral cortex. The algorithm is dependent on thalamic nuclei priors and local fiber orientation. Thalamus connectivity-based parcellation methods is used for the validation of the algorithm. Algorithm successfully segmented the anatomical plausible thalamic nuclei.
Li et al. \cite{li2018structure} segmented hepatocellular carcinoma (HCC) nuclei by using structure convolutional extreme learning machine (SC-ELM) and case-based shape template (CBST). First pathology images are globally segmented where each connected region is treated as nucleus clump. Then contour refinement of nucleus clumps is done through a probability model of three energy function. And at last, a combination of CBST method and pixel-based classification is used for the obscure boundary inference. This method, evaluated on 127 liver pathology images shows a good result in comparison with other related work.
 Song et al. \cite{decisiontrees1} designed a representation learning decision trees based sparse coding ensembles network where fast Intra-decoders and Inter-Encoder is used for enhancing connectivity patterns as well as iterative regression, via decision tree based  ensemble mappings through regularization, pruning and random sampling for increasing model generalization. 

Jie et al. \cite{decisiontrees2} proposed an improved multi layer sparse convolution model (ML-BSC) where handcrafted approach is utilized for robust feature extraction and optimized computation via integrating discriminative probability based decision tree ensemble for enhancing classification performance. 
Luna et al. \cite{siamese} addressed the problem of pinpoint boundary delineation of adjacent nuclei by introducing a novel deep neural network. The network makes the prediction about the instances that whether they from individual grouped nuclei. It uses the decision making with Siamese network for learning the relation between the two adjacent instances and surrounding features of the adjacent nuclei. The network further predicts the class and their overlapping dice score through a decoding network improving classification accuracy. The network exhibit significant improvement in cell separation accuracy. 
A joint contour based boundary extraction method is used by Kurmi et al. \cite{cannycontour} via 3 stage cascaded network for slides pre-processing, nuclei points extraction  and region refining via canny edge detector, and composite nuclei segmentation through contour estimation.
Rashmi et al. \cite{colornorm} developed an unsupervised model for segmenting the nuclei from breast histopathological images based on Chan-Vese model. It pre-processes images via color normalization for discriminating foreground instances and background, followed by multi-channel learning based on color features for efficient segmentation. Song et al. \cite{7339670} used a combination of  watershed and GVF snake model for nuclei separation from the background followed by Convex hull detection and concave point detection for splitting of overlapped nuclei and yielded 91\% precise results \cite{concavepoint1}. Lapierre-Landry et al. \cite{watershednew} proposed a regression based joint V-Net and watershed based approach in 2021 for 3-D microscopy nuclei segmentation on embryonic heart dataset having high cell density, with an accuracy of segmenting 1000 nuclei centroids in under a minute. In SEENS \cite{Zhao2021SEENSNS} mathematical morphology operators are integrated with selective search for segmenting nuclei from cervical images while eliminating non-nuclei regions and avoiding repeated segmentation. Canny edge operator is used for extracting edge information for enhancing nuclei edge selection precision. Ramirez et al. \cite{9630846} focused on morphological transformations and adaptive intensity adjustments for segmentation.
 Karthick et al. \cite{8857080} used wavelet decomposition technique followed by random forest classifier for nuclei segmentation on thalamus dataset and further post processed segmentation output for refinement reporting 75\% Accurate segmentation results.
 
Roy et al. \cite{challenge5} proposed an energy maximization function for nuclei instance segmentation. They segmented cell nuclei via contrast adaptive technique followed by intensity adaptive weighted thresholding and energy maximization segmentation on Kumar dataset resulting in 87\% dice accuracy.

\paragraph{\textbf{Stand out Method}}
Among all hand crafted feature extraction method Hameed et al. \cite{watershednew3}, a global entropy thresholding-based segmentation technique, outperformed all other state-of-art hand crafted feature extraction methods for nuclei instance segmentation by achieving MAD of 0.478, DC of 0.967 and accuracy of 0.970.

\begin{tiny}
\begin{longtable}{ |m{0.5cm}| m{1.8cm} | m{1cm} | m{2cm}  | m{2cm} | m{2cm} | m{1.5cm}|}
 
  \caption{Hand Crafted featured based nuclei instance segmentation strategies} \label{table4} \\
   
  \hline
    \rowcolor[HTML]{FBEBD3}
    \textbf{S.No} &
    \textbf{Reference} &
    \textbf{Year}&
    \textbf{Organ}&
    \textbf{Dataset} &
    \textbf{Feature Extraction Method} &
    \textbf{Performance}\\
\hline
\endfirsthead

\multicolumn{3}{c}%
{{ }} \\
\hline 

\endhead

\hline 

\endfoot

\hline 
\endlastfoot
\rowcolor[HTML]{fdf3e7}
1 &     Rojas-Moraleda et al. \cite{rojas2017robust} & 2017 & Liver & Liver dataset	& Topological features based approach & 51\% F1 Score \\  
\hline
    \rowcolor[HTML]{fdf3e7}
2 &    Gautam et al. \cite{gautam2017unsupervised} & 2017 & Cervix & Herlev dataset	& SLIC, mean shift clustering, adaptive thresholding & 70\% Accuracy \\  
\hline

  \rowcolor[HTML]{fdf3e7}
3 &    Reljin et al. \cite{reljin2017multifractal} & 2017 & Fish images & Institute of Pathology, University of Bern, Switzerland	& Matrix of holders exponent thresholding & 90\% Accuracy \\
 
\hline

  \rowcolor[HTML]{fdf3e7}
4 &    Brieu et al. \cite{brieu2017learning} & 2017 & Breast & 30 H\&E images	& Local maxima detection & 75\% F1 Score \\
\hline
\rowcolor[HTML]{fdf3e7}
5 & Xu et al.\cite{xu2016automatic} & 2017 & Skin, Breast & Skin histopathological images dataset	& gLoG kernel, mean shift clustering & 91\% F1 Score \\  
\hline


  \hline

 \rowcolor[HTML]{fdf3e7}
6 &  Battistella et al.\cite{Battistella2016RobustTN} & 2017 & Brain thalamus & Thalamic Brain Anatomy dataset	&  ODFs, K-Means clustering  & 84\% Dice Score\\

\hline
\rowcolor[HTML]{fdf3e7}
7 & Hameed et al. \cite{watershednew3} & 2017 & Breast & OSCC & Entropy, Thresholding, Watershed, GLCM & 96\% DSc    \\
\hline
\rowcolor[HTML]{fdf3e7}
8 & Song et al. \cite{7339670} & 2017 & Kidney & Farsight dataset, TCGA & LoG filter, Morphogical operation, Regression model  & 70\% DSc    \\
\hline
\rowcolor[HTML]{fdf3e7}
9 & Saha et al. \cite{saha2018segmentation} & 2018 & Cervix & ISBI-2014 Cervical cytology dataset & LoG filter, Morphological operation, Regression model  & 95\% DSc    \\

\hline
    \rowcolor[HTML]{fdf3e7}
10 & Lee et al. \cite{lee2017nucleus} & 2018 & breast & BBBC006v1, U20S & LoG filter, Thresholding, Ellipse fitting  & 95\% DSc    \\
\hline
    \rowcolor[HTML]{fdf3e7}
11 & Kostrykin et al. \cite{challenge6} & 2018 & Flourescence microscopy images & NIH3T3, ISBI-2013    & Second order optimization & 75\% F1- Score   \\
\hline

\rowcolor[HTML]{fdf3e7}
12 & Li et al. \cite{li2018structure} & 2018 & Liver & Hepatocellular carcinoma (HCC) & Mean     shift clustering & 92\% Accuracy  \\
 \hline
   \rowcolor[HTML]{fdf3e7}
13 & Quachtran et al. \cite{count} & 2018 & Flourescene confocal microscopy images & CLARITY Dataset & Iterative radial voting & 64\% Dice Score  \\
 \hline
   \rowcolor[HTML]{fdf3e7}
14 & Semedo et al. \cite{semedo2018thalamic} & 2018 & Thalamus & Thalamic nuclei dataset & Local fibre orientation segmentation & 70\% F1- Score  \\
 \hline
  \rowcolor[HTML]{fdf3e7}
15 & Guo et al. \cite{guo2018clumped} & 2018 & breast & Flourscene Microscopy Dataset & Curvature point detection, Ellipse fitting &  69\% Accuracy \\
\hline

  \rowcolor[HTML]{fdf3e7}
16 & Karthick et al.\cite{8857080} & 2019 & Thalamus, Brain  & Private dataset	&  Wavelet decomposition, Random forest classifier & 75\% Accuracy \\  
\hline
     \rowcolor[HTML]{fdf3e7}
17 & Song et al.\cite{decisiontrees2} & 2019 & Breast, Prostate, Kidney, Liver, Stomach, Bladder
& Kumar, KIRC, Farsight	& Decision trees & 78\% Dice Score \\

 \hline
 \rowcolor[HTML]{fdf3e7}
18  & Song et al.\cite{decisiontrees1} & 2021 & Cervical cytology & ISBI-2015 &  Depth first search strategy, Decision trees  &  95\% F1 Score\\

  \hline
   \rowcolor[HTML]{fdf3e7}
19 & Dongyao et al.\cite{concavepoint1} & 2021 & Breast, Cervix & BJTU BIT, U20S NIH3T3  & Watershed, GVF snake model, Ellipse fitting, Convex detection  & 87\% Dice Score\\
 
\hline
\rowcolor[HTML]{fdf3e7}
20 & Zhao et al. \cite{Zhao2021SEENSNS} & 2021 & Cervix & 18 Whole slide cervical cell images & Selective search, Canny edge, Morphological operation & 92\% Dice Score  \\
 \hline
  \rowcolor[HTML]{fdf3e7}
21  & Ramirez et al. \cite{9630846} & 2021 & Breast 
& BreakHis database
& Morphological transformation, Adaptive intensity adjustment
& 78\% Dice Score \\

\hline
\rowcolor[HTML]{fdf3e7}
22 & Roy et al. \cite{challenge5} & 2021 & Bladder, Stomach, Colorectal, Breast, Kidney, Liver, Prostate

& Kumar, KMC MAHE (80 H\&E stained images)

& Thresholding, Energy maximization function

& 87\% Dice Score \\
   
  \end{longtable}
 \end{tiny}
 
\subsubsection{\textbf{Deep Learning based methods}}
\paragraph{\textbf{ Region based Nuclei localization }}

Regional proposal based segmentation approach was initially designed for natural images segmentation. However, in last few years they have been proved extremely influential for other domains as well and are highly adapted for medical images segmentation too specially in  nuclei and gland segmentation tasks. Main idea behind proposal base architecture is detecting regions according to the variations in similarity metrics and color differences, followed by classification for regions having high probability of object existence, they are often regarded as region wise prediction as well. Mask base region proposal network also called as Mask-RCNN comprising of CNN as base for feature extraction and region proposal network for suggestion about object regions which are further used for prediction of binary masks. Mask-RCNN along with U-Net has been widely used in various nuclei segmentation tasks in past few years. Table \ref{table5} is showing the details of the articles reviewed.
In 2018 Liu et al. \cite{rpn3} combined (Mask-RCNN) and fully connected conditional random field (LFCCRF) for segmenting cervical nuclei via localizing nuclei boundaries through multi-scale feature maps through regional proposals followed by enhancing segmentation by passing spatial information to LFCCRF for further refinement. However, accuracy of this model required further enhancement. 
Voula et al. \cite{vuola2019maskrcnn} comparatively analysed both regional proposal and U-Net segmentation techniques and ensembled architecture. For circular shaped medium sized nuclei Mask-RCNN model gave excellent results but its performance degraded in case of elliptical shapes however ensembled model resulted in overall accuracy enhancement.

Baykal et al. \cite{rpn5} used object detection models including Faster-RCNN, region based convolution neural network and Single Shot Detection model (SSD) for the first time for pathology image nuclei detection. However, SSD results for detection were not up to the mark while Faster-RCNN along with ResNet yielded best results. A similarity learning based approach is used by Sun et al. \cite{rpn6}, they introduced an embedding layer for building networks and training through embedding loss function. These networks were able to learn distinguishing features on the basis of similarity score which is further used for instance classification. A unique module incorporating guided anchoring into regional proposals is used in \cite{rpn4} for candidate proposals generation along with a new branch for regressing intersection over union(IoU) between ground truths and detection boxes for bounding box localization. They also passed FBS to soft non maxima suppression  for true positive box preservation.
\paragraph{\textbf{Stand out Method}}
The solution of Liang et al. \cite{rpn2} was found significantly better than all other previous state of the art region proposal based patch extraction techniques for instance segmentation of nuclei and glands. It yielded best performance results for generalization and other metrices on two major publicly available datasets including DSB-2018 and MoNuSeg.
\begin{tiny}
\begin{longtable}{ |m{0.5cm}| m{1.5cm} | m{1cm} | m{2.3cm}  | m{2cm} | m{2cm} | m{1.5cm}|}
 
  \caption{Nuclei Instance Segmentation via two stage Patch Extraction techniques} \label{table5} \\
   
  \hline
    \rowcolor[HTML]{FBEBD3}
    \textbf{S.No} &
    \textbf{Reference} &
    \textbf{Year}&
    \textbf{Organ}&
    \textbf{Dataset} &
    \textbf{Technique} &
    \textbf{Performance}\\
\hline
\endfirsthead

\multicolumn{3}{c}%
{{ }} \\
\hline 

\endhead

\hline 

\endfoot

\hline 
\endlastfoot
\rowcolor[HTML]{fdf3e7}
1 & Liu et al.\cite{rpn3} & 2018 & Pap smear images & Herlev, BNS, MOD &  Mask-RCNN, CRF & 84\% F1 \\   

 \hline
 \rowcolor[HTML]{fdf3e7}
2 & Naylor et al.\cite{rpn1} & 2019 & Bladder, Stomach, Colorectal, Breast, Kidney, Prostate & IIT	&  U-Net, FCN, Mask-RCNN & 81\% F1 \\ 
\hline
\rowcolor[HTML]{fdf3e7}
3 & Sun et al.\cite{ rpn6} & 2019 & Breast, Kidney, Liver, Prostate, Bladder, Colon & MoNuSeg &  R-CNN, ResNet & 85\% F1 \\
 \hline
\rowcolor[HTML]{fdf3e7}
4 & Feng et al.\cite{ rpn4} & 2019 & Breast, Kidney, Liver, Prostate,  Bladder, Colon & DAPI, TNBC & Mask-RCNN & 0.54 AJI \\
 \hline
 \rowcolor[HTML]{fdf3e7}
5 & Vuola et al. \cite{vuola2019maskrcnn} & 2019 & Colon & Fluorescence Images & Mask-RCNN, U-Net & 72\% DSc \\
\hline
\rowcolor[HTML]{fdf3e7}
6 & Jung et al. \cite{Jung2019AnAN} & 2019 & Breast, Kidney & BNS, MOD & Mask-RCNN
& 86\% F1 \\
 \hline
\rowcolor[HTML]{fdf3e7}
7 & BayKal et al.\cite{rpn5} & 2020 & Lungs & Pleural Effusion Cytology Images 	& Faster- RCNN, SSD & 98\% F1 \\
\hline
\rowcolor[HTML]{fdf3e7}
\textbf{8} & \textbf{Liang et al.}\cite{rpn2} & \textbf{2022} & \textbf{Breast, Kidney, Lung, Prostate, Bladder, Colon} & \textbf{DSB-2018, MoNuSeg} &  \textbf{Mask-RCNN, FPN} & \textbf{84\% F1} \\
  \end{longtable}
 \end{tiny}
\paragraph{\textbf{Encoder Decoder based segmentation techniques}}

Convolutional neural networks have always been the most used techniques for vision problems and have achieved great success in detection and segmentation tasks in medical image analysis. However, in deep neural network based learning it was observed that CNNs despite having the ability of learning major features via layer wise propagation at times results in lost spatial information.
 Compared to simple deep neural networks, fully convolutional neural networks (FCN) based  nuclei segmentation frameworks were found more efficient in  nuclei and gland segmentation in histopathology images. For instance, while classical CNNs works by classifying individual pixels via sliding window approach, FCN allows up scaling  classical features hence segmenting images in single pass. For resolving shortcomings of CNN during segmentation, encoder decoder based approaches were proposed. It works on the principle of down sampling first for better feature learning followed by upsampling for final segmentation. In this architecture during down sampling network reduces activations size through feed forward neural network, while in upsampling deconvolution network or unpooling operations are primarily used for regaining lost information. The table \ref{table6} summarizes details about encoder decoder based segmentation techniques.
In 2017, Zhang et al. \cite{7950548} used FCN for segmenting whole cell into cytoplasm, background and nuclei probabilistic map. It captured nuclei deep features precisely however accurate boundaries localization for segmented results was still an issue. As a resolution, graph based segmentation is incorporated on FCN segmented features thus improving nucleus boundaries detection via probability map. Similar encoder decoder architecture for nuclei instances segmentation is proposed by Graham et al. \cite{8363645}. They used down sampling operation for strong gradient features while in back propagation residual blocks and max pooling operators are used, similarly for upsampling deconvolution operators were incorporated for fine image rebuilding. A new stain aware auxiliary loss function resulted in better performance by targeting only low haematoxylin intensity based areas within nuclei and high intensity areas in background. Xu et al. \cite{8759530} proposed a joint segmentation detection module using U-Net and SSD for reducing segmentation inaccuracies of previous approaches. It dynamically integrated nuclei location output and semantic segmentation results for joint performance boost. A deep regression distance based approach by Naylor et al. \cite{rpn1} resolved the segmentation issues faced for closed touching nuclei. In this architecture pre-trained weights of VGG-16 and training results of three different models including FCN, Mask-RCNN and U-Net were compared for prediction of unnormalized intra nuclear distance map. Despite having up to the mark accuracy, high computational complexity and generalization inability were still a major challenge of this approach.
On the other hand stage wise segmentation approach was used by Qu et al. \cite{unet8} and Zhao et al. \cite{bayseian2}. Two similar SU-Net architectures were utilized in this approach  where first stage aimed at segmenting nuclei regions, while second stage was designed for segment overlapping nuclei regions. Finally for instance segmentation results output of both stages were merged together. This multi stage learning approach outperformed all previously existing state of the art approaches.

A shape prior regularized network architecture is proposed by Tofighi et al. \cite{regularization}. In this technique initial learnable layers, learns from prior information (generated edge map through raw input image and predefined shapes) via fixed processing and performs nuclei detection consistent to boundary.

  Cui et al. \cite{onestepcontour2019} proposed a model for simultaneously predicting nuclei and their contours at the same time via a nuclei boundary prediction model using  attention based segmentation and fast non parameter dilation based post-processing after contour based segmentation via fully convolutional neural network.

 Narotamo et al. \cite{unet13} devised a model that initially divides image into equal sized overlapping patches and segments nuclei and boundary map by fully convolutional neural network (FCN). Detecting nuclei boundary in each patch enabled splitting touched and overlapped nuclei thus improving accuracy.  
 Han et al. \cite {gan6} focused on unsupervised learning approach via generating 3D nuclei data using Recycle-GAN along with Hausdorff distance loss for nuclei shape preservation. For segmentation and classification of synthetically generated data a 3D CNN is employed. Similar, 3D nuclei segmentation method has been proposed by Ho et al. \cite{unet17, 3d1} and Guan et al. \cite{unet18} via distance transform, adaptive histogram equalization, and a 3-D convolution neural network for classification CNN by searching nuclei centers and spatial \& channel attention module for nuclei segmentation is proposed in 3D encoder decoder model trained on initially generated synthetic volumes. 
 An enhanced approach for resolving over fitting issue faced during multi resolution feature extraction of conventional U-Net and enhancing convergence performance Majdi et al. \cite{resunet1}, \cite{resu2} and Xie et al. \cite{unet7} proposed a residual U-Net based technique for segmentation where they incorporated batch normalization and drop out layers in architecture with the addition of scale-wise triplet learning and count ranking for vanishing gradient and exploding loss risks mitigation during segmentation. Mandal et al. \cite{unetnew4} proposed a Y shaped model where a novel forked decoder is tied to segmentation and deblurring, this additional element in model resulted in fine tuning output for normal as well as blurred out of focus images.
Similar technique is proposed by Aatresh et al. \cite{attention3} with a slight enhancement of dimension wise convolution combined with atrous spatial pyramid pooling for improved encoder decoder architecture and multi task learning based on nuclei region and boundaries extraction. Similar architecture is proposed by Pan et al. \cite{unet3} and Dabass et al. \cite{gland16} where along with features learning, diminishing gradient issue is alleviated as well. For resolving problems faced by model during small objects features capturing a dual encoder architecture is proposed by Narotamo et al. \cite{unet13}. They utilized prior features information in encoding network and input feature maps are utilized in  attention skip modules for segmentation performance boost. Vahadane et al. \cite{inproceedings} used  post processing approach for separating touching nuclei and semantic segmentation of nuclei. Firstly, objects are thresholded followed by boundary map substraction for nuclei instance map segregation, then instance based energy map is generated for pixel to background distance calculation. Erosion is used for marker generation which are then passed to watershed along with distance map for final processing.

 Sparse reconstruction based deconvolution technique is used in \cite{dcn1, dcn2, dcn3, UNET4} and \cite{unetnew}. They used dilated dense block with exponential increase in dilation rate for encoding cascaded multi stage neural networks information which is further trained via gradient descent technique and attention score map is used here as regularization factor thus avoiding binding and interposed nuclei overlaps. Image level and instance level alignment based on domain shift minimization approach is proposed by Wang et al. \cite{unet19} where INA initially extracted instance features through nuclei locations via a temporal ensembling based Nuclei Localization (TENL) module, this resulted in automatic candidate nuclei location generation. Huang et al.\cite{unet20} proposed a shared decoder path instead of conventional two path decoder technique used by Qingbo et al. \cite{unetnew3} thus increasing recognizability range generated through new half path served as a natural proxy for curriculum learning model. Kablan et al.\cite{segnet1}  designed a SegNet based architecture, which is one of the best model for problems dealing with image segmentation tasks due to its property of direct information transfer instead of convolution. 
 
 For accurately segmenting overlapping and cluttered nuclei Wang et al. \cite{wang2020bending} proposed a bending loss regularized network using multi task  learning approach inspired from HoVer-Net architecture by Graham et al. \cite{graham2019hover}. It introduced generalization by training decoders for 3 streams including nuclei distance map prediction, overlapped nuclei distance map and instance segmentation via using high penalty bending loss for large curvature contour points and low penalty for small curvature for simultaneous concave or convex transformation. While in HoVer-Net 3 branches were nuclei pixel branch (for pixels separation from background), hover branch in which nuclei horizontal and vertical distances with respect to center of masses is used for separating clustered nuclei thus resulting in accurate segmentation specially for cases having large number of overlapping nuclei segmented instances and lastly nuclei classification  where  segmented instances were further passed to a dedicated upsampling branch for type classification. HoVer-Net was also the first ever technique achieving both segmentation and classification via same network. Zhao et al. \cite{ZHAO2020101786} also used similar 3 branch architecture having RGB branch, segmentation branch and haematoxylin branch simultaneously. In this technique RGB branch was used for raw features extraction for segmentation task, H branch for H\&E aware feature extraction for the task of nuclei contour detection task and segmentation branch fused both RGB and Haematoxylin contour features for final results prediction. Braga et al. \cite{contour1} also used contour based boundaries detection and encoder decoder based learning approach.

 Li et al. \cite{8744511} improved shortcomings of previous regression based segmentation and devised a multi task learning fusion based approach for improving glioma nuclei segmentation accuracy via boundary and region information they adopted a U-Net based encoder decoder where paired upsampling paths are used for boundary classification and touching nuclei separation while in other path regression model is used for nuclei region distance map prediction approach. These layers are further fused together for final segmentation. In 2017 Fu et al. \cite{watershednew4} utilized  watershed for cell nuclei quantification in parallel with CNN based segmentation. Luna et al. \cite{siamese} used an encoder decoder based approach where contextual information is captured via harnessing multi level CNN capabilities and features concatenation along with up sampled features using skip connections followed by post-processing via morphological operations. This model detected malignant cases accurately but failed in case of some benign test samples due to erosion during noise removal process. A modification to this approach was proposed in region proposal and encoder decoder based joint nuclei detection and segmentation technique given by Cheng et al. \cite{9180333} where both shallow and deep layers are merged  via skip  connections for accurate detection of nuclei locations using bounding boxes and detected patches are further passed to U-Net for segmentation. Voula et al. \cite{vuola2019maskrcnn} gave a comparative analysis of both regional proposal and U-Net model based nuclei segmentation as well as ensembled architecture. For circular shaped medium sized nuclei, Mask-RCNN model gave excellent results but its performance degraded in case of elliptical shapes. It had some trouble in approximating bounding boxes for the nuclei while U-Net performed better in such cases, similarly it was also found that an ensembled model resulted in overall accuracy enhancement. 
 A novel kernalized correlation filters are used for nuclei detection in  \cite{correlation1} due to its dependence on small training dataset as well as interpretable and computationally efficient nature as compared to other deep learning based models and can give state of the art results. Similar approach is used in \cite{JAVED2021102104} where spatial structure is incorporated via constructing a dense graph based on different deep features across different nuclei components then correlation filter is used for discriminating nuclear and non-nuclear region.
 Similarly, Salvi et al. \cite{watershedlatest} used watershed segmentation in multi scale framework after progressive weighted mean detection for object detection and area based correction for aligning over or unsegmented objects. In this technique watershed transform is specifically used for nuclei detection purpose. In \cite{watershed1, watershed2} color de-convolution is used for image preprocessing for H\&E stain highlighting which are then trained on deep CNN followed by marker controlled seeded watershed technique for splitting touching nuclei but this process specially semantic segmentation used in CNN results in a computationally expensive model and performance degraded in case of highly overlapped nuclei further resulted in over segmentation as well.
 An ensembling based improved architecture is proposed by Zhao et al. \cite{8759262} and Liu et al. \cite{unet2} having a U shaped ensembled convolution network as backbone with dense blocks for effectively transferring features information alongside overcoming vanishing gradient problem of prior architectures. Another feature added in this model was deformable convolution for dealing with nuclei of different sizes and irregular shapes thus enhancing model flexibility. A point annotation based architecture is presented by Yoo et al. \cite{point1} as PseudoEdgeNet. In this a guided segmentation network is trained for recognizing nuclei edges without prior annotations through multi scale pyramidal model and backbone ResNet in Kanwal et al. \cite{unet16}. Another similar one click approach for quick annotation collection is proposed by koohbanani et al. \cite{annotation} where they used a single click for precise annotation for single as well as for structures comprising of multiple nuclei such as glands. In 2021, Valkonen et al. \cite{transferlearning} proposed a supervised transfer learning model via applying unsupervised domain adaptation for model generalization on seen as well as unseen labeled images data and achieved 77\% accuracy. Another transfer learning and logistic regression based approaches is used by Li et al. \cite{transferlearning1} via unsupervised sparse auto encoder (SSAE) and case based post processing module (CPM) technique. They extracted high level features via transfer learning followed by logistic regression classifier(LRC) on extracted features and fine tuning via CPM.
 
 \paragraph{\textbf{Stand out Method}}
Internal co-variance shift based U-Net architecture proposed by Wan et al. \cite{unet15} in 2021, stood out as a top performing model due to its generalized network across diverse set of images, experimental inputs variation for a heavily imbalanced dataset. With 80\% of training images it yielded an accuracy of 95\%. Similarly, FPN based encoder decoder architecture proposed by Yang et al. \cite{9098383} outperformed previous state of the art result with 97\% F1-Score.

\begin{tiny}
\begin{longtable}{| m{0.4cm}| m{1.2cm} | m{0.5cm} | m{2cm} | m{2.2cm} | m{2.2cm} | m{2.5cm}|}
 
  \caption{Encoder Decoder based Segmentation Methods} \label{table6} \\
   
  \hline
    \rowcolor[HTML]{FBEBD3}
\textbf{S.No}&
\textbf{Reference}&
\textbf{Year}&
\textbf{Organ}&
\textbf{Dataset}&
\textbf{Technique}& 
\textbf{Performance} \\ 
\hline
\endfirsthead

\multicolumn{3}{c}%
{{ }} \\
\hline 

\endhead

\hline 

\endfoot

\hline 
\endlastfoot
  
     \rowcolor[HTML]{fdf3e7}
      1 & Zhang et al.\cite{7950548} & 2017 & Cervix & Herlev & FCN  & 92\% Zijdenbos similarity  \\
    \hline
     \rowcolor[HTML]{fdf3e7}
        2 & Ho et al.\cite{unet17} & 2017 & Kidney & Rat kidney dataset & FCN  & 92\% Accuracy \\
    \hline
     \rowcolor[HTML]{fdf3e7}
      3 & Pan et al.\cite{dcn1} & 2017 & Breast & David Rimm's dataset & FCN  & 83\% F1 Score \\
    \hline
        \rowcolor[HTML]{fdf3e7}
    4 & Fu et al. \cite{watershednew4} & 2017 & Kidney & Rat kidney dataset & SegNet, Watershed & 94\% mAP  \\
    \hline
   \rowcolor[HTML]{fdf3e7}
    5 & Graham et al.\cite{8363645} & 2018 & Breast & CPM-17 & SAMS-NET & 80\% Dice Score \\
       \hline
   \rowcolor[HTML]{fdf3e7}
    6 & Ho et al.\cite{3d1} & 2018 & Kidney & Rat Kidney dataset & CNN, seed candidate selection &  94\% F1 Score \\
           \hline
   \rowcolor[HTML]{fdf3e7}
    7 & Hofener et al.\cite{9012} & 2018 & Breast, Colon & Warwick-QU & FCN, Pmap & 82\% F1 Score \\
    
    \hline
     \rowcolor[HTML]{fdf3e7}
    8 & Naylor et al.\cite{dcn2} & 2017 & Breast, kidney & U20S & FCN \& PangNet & 80\% F1 Score \\
    \hline
     \rowcolor[HTML]{fdf3e7}
        9 & Zhao et al.\cite{8759262} & 2019 & cervix & Herlev & U-conv + deformable conv  & 93\% Zijdenbos similarity \\
    \hline
       \rowcolor[HTML]{fdf3e7}
        10 & Zhao et al.\cite{bayseian2} & 2019 & Breast & University Hospital Poland & FCN, Bayesian Inference  & 83\% accuracy \\
    \hline
     \rowcolor[HTML]{fdf3e7}
        11 & Tofighi et al.\cite{regularization} & 2019 & Cervix & Herlev & U-conv + Deformable convolution  & 93\% Zijdenbos similarity \\
    \hline
     \rowcolor[HTML]{fdf3e7}
    12 & Kingbo et al.\cite{unetnew3} & 2019 & multiple & Kumar, TNBC & Stacked U-Net & 80\% F1 Score \\
    \hline
     \rowcolor[HTML]{fdf3e7}
    13 & Lee et al.\cite{UNET1} & 2019 & Breast & Fluorescence Image Dataset &  OS-NET  & 70\% Dice Score \\
    \hline
     \rowcolor[HTML]{fdf3e7}
    14 & Zeng et al.\cite{resu2} & 2019 & multiple & TCGA &  Residual Inception U-Net & 77\% F1 Score \\
    \hline
     \rowcolor[HTML]{fdf3e7}
    15 & Voula et al.\cite{vuola2019maskrcnn} & 2019 & Colon & UW dataset & SP-CNN & 85\% F1 Score \\
    \hline
     \rowcolor[HTML]{fdf3e7}
      16 & Cui et al.\cite{onestepcontour2019} & 2019 & multiple & MOD, BCD & FCN & 84\% F1 Score \\
    \hline
       \rowcolor[HTML]{fdf3e7}
    17 & Xu et al.\cite{ 8759530} & 2019 & Bladder, Stomach, Breast, Kidney, Liver, Prostate & MoNuSeg & SSD + U-Net & 80\% Precision Accuracy \\
    \hline
     \rowcolor[HTML]{fdf3e7}
    18 & Navid et al.\cite{annotation} & 2019 & Bladder, Colon, Stomach, Breast, Kidney, Liver, Prostate & GlaS, CRAG, MoNuSeg & NuClick & 91\% Dice Score \\
    \hline
  
     \rowcolor[HTML]{fdf3e7}
    19 & Li et al.\cite{8744511} & 2019 & multiple & Glioma, MoNuSeg &  U-Net, Distance Map, Fusion Module & 76\% F1 Score \\
    \hline
     \rowcolor[HTML]{fdf3e7}
    20 & Pan et al.\cite{unet3} & 2019 & Bladder, Colon, Stomach, Breast, Kidney, Liver, Prostate & BNS, MOD &  ASU-Net & 87\% F1 Score  \\
    \hline
     \rowcolor[HTML]{fdf3e7}
    21 & Yoo et al.\cite{point1} & 2019 & Bladder, Stomach, Breast, Kidney, Liver, Prostate & MoNuSeg, TNBC &  FPN, ResNet-50 & 60\% Accuracy  \\
  \hline
   \rowcolor[HTML]{fdf3e7}
    22 &  Graham et al.\cite{graham2019hover} & 2019 & Bladder, Stomach, Breast, Kidney, Liver, Prostate & CoNSeP, Kumar	&  U-Net
  & 80\% Dice Score 
  \\
  \hline
   \rowcolor[HTML]{fdf3e7}
    23 & Qu et al.\cite{UNET4} & 2019 & Lungs & Lung Cancer Dataset	& FCN + U-Net & 88\% F1 Score \\  
    \hline
       \rowcolor[HTML]{fdf3e7}
    24 & Liu et al.\cite{unet2} & 2019 & Bladder, Stomach, Breast, Kidney, Liver, Prostate & Kumar, TNBC &  Stacked  U-Net & 80\% F1 Score \\
    \hline
     \rowcolor[HTML]{fdf3e7}
    25 & Long et al.\cite{unetnew} & 2020 & Bladder, Stomach, Breast, Kidney, Liver, Prostate & DSB-2018 & U-Net \& & 62\% Accuracy \\
    \hline
   
   \rowcolor[HTML]{fdf3e7}
    26 & Zhao et al.\cite{ ZHAO2020101786} & 2020 & Bladder, Stomach, Breast, Kidney, Colorectal adenocarcinomas, Prostate & MoNuSeg, CoNSeP, CPM  & U-Net \& & 83\% Dice Score \\
    \hline
     \rowcolor[HTML]{fdf3e7}
      27 &  Kowal et al.\cite{watershed2} & 2020 & Breast & David Rimms dataset	& FCN + watershed & 90\% Accuracy \\
      
    \hline
     \rowcolor[HTML]{fdf3e7}
     28 &  Qu et al.\cite{bayesian1} & 2020 & Bladder, Stomach, Breast, Kidney, Liver, Prostate & Lung cancer dataset, MOD	& Bayesian CNN & 83\% F1 Score \\
    \hline
     \rowcolor[HTML]{fdf3e7}
      29  & Cheng et al.\cite{ 9180333} & 2020 & Bladder, Stomach, Breast, Kidney, Liver, Prostate & DSB-2018  & UNet, FPN \& & 77\% Accuracy \\
  \hline
   \rowcolor[HTML]{fdf3e7}
  30 & Xie et al.\cite{unet7} & 2020 & Bladder, Stomach, Breast, Kidney, Liver, Prostate & MoNuSeg	& Res-UNet & 65\% accuracy \\   
  \hline
   \rowcolor[HTML]{fdf3e7}
  31 & Qu et al.\cite{unet8} & 2020 & Lungs, Prostate & Lung Cancer,	&  U-Net, K-Means & 75\% F1 Score \\
  \hline
  \rowcolor[HTML]{fdf3e7}
  \textbf{32} & \textbf{Yang et al.}\cite{9098383} & \textbf{2020} & \textbf{Cervix} & \textbf{Herlev} & \textbf{FPN}, \textbf{Encoder decoder} & \textbf{97\% F1 Score} \\
  \hline
   \rowcolor[HTML]{fdf3e7}
  33 & Baykal et al.\cite{segnet1} & 2020 & Breast, Lung & Lung Cancer Dataset, U20S & FCN, SegNet, U-Net & 95\% Dice Score \\
  \hline
    \rowcolor[HTML]{fdf3e7}
    34 & Han et al. \cite {gan6} & 2020    & Kidney  & Kidney dataset & U-Net, Recycle-GAN   & 82\% F1 Score                                       \\
  \hline
     \rowcolor[HTML]{fdf3e7}
      35 & Hussain et al.\cite{challenge4} & 2020 & breast & Smear dataset & FCN & 96\% ZSI \\
    \hline
    \hline
 \rowcolor[HTML]{fdf3e7}
36 &  Xei et al.\cite{watershed1} & 2020 & Breast, Prostate, Kidney, Stomach & Kumar, MICCAI-2017 &  Marker controlled watershed & 87\% F1 Score \\

   \rowcolor[HTML]{fdf3e7}
  37 & Narotam et al.\cite{unet13} & 2021 & Retina & Mouse-Retina Dataset	&  U-Net & 63\%  Score \\
  
   \hline
   \rowcolor[HTML]{fdf3e7}
  38 & Huang et al.\cite{unet20} & 2021 & Deep sea archea & Fluorescence microscopy images	&  Encoder decoder  & 70\%  F1 Score \\
  \hline
   \rowcolor[HTML]{fdf3e7}
   
      39 & Vahadane et al.\cite{inproceedings} & 2021 & Bladder, Stomach, Breast, Kidney, Liver, Prostate & CPM-17, Kumar, CoNSeP	&  U-Net, Attention skip module & 81\%  Dice Score \\
  \hline
   \rowcolor[HTML]{fdf3e7}
  
 \textbf{40} & \textbf{Wan et al.}\cite{ unet15} & \textbf{2021} & \textbf{Bladder, Stomach, Breast, Kidney, Liver, Prostate} & \textbf{MoNuSeg, TNBC} & \textbf{U-Net, Internal, Co-variance Shift} & \textbf{95\% F1 Score} \\
  \hline
   \rowcolor[HTML]{fdf3e7}
    41 & Kim et al.\cite{dcn3} & 2020 & Brain & Diffusion weighted MRIs & DCN-Net & 87\% Dice Score \\
  \hline
   \rowcolor[HTML]{fdf3e7}
    42 & Wang et al.\cite{wang2020bending} & 2020 & Bladder, Stomach, Breast, Kidney, Liver, Prostate & MoNuSeg & Bending loss, U-Net & 83\% Dice Score \\

  \hline
   \rowcolor[HTML]{fdf3e7}
     43 & Valkonen et al.\cite{transferlearning} & 2021 & Bladder, Stomach, Breast, Kidney, Liver, Prostate &  MoNuSeg & VGG, Transfer learning, FCN & 77\% F1 Score \\
  \hline
   \rowcolor[HTML]{fdf3e7}
    44 & Mandal et al.\cite{unetnew4} & 2021 & Bladder, Stomach, Breast, Kidney, Liver, Prostate & Fluorescence image dataset & U2OS & 84\% F1 Score \\
  \hline
  \rowcolor[HTML]{fdf3e7}
    45 & Maryse et al. \cite{watershednew} & 2021 & Heart & Heart Dataset & V-NET, Centroid calculation & 95\% F1    \\
    \hline
   \rowcolor[HTML]{fdf3e7}
      46  & Braga et al.\cite{contour1} & 2021 & Cervical cytology images & ISBI-2014, Herlev	&  Multi scale narrow band level set algorithm & 85\% Dice Score \\

         \hline
   \rowcolor[HTML]{fdf3e7}
 47 &  Kanwal et al.\cite{unet16} & 2022 & Bladder, Stomach, Breast, Kidney, Liver, Prostate & MoNuSeg, TNBC, CryoNuSeg	& Distance map, Skip Connection, U-Net & 90\% F1 Score \\
  \hline
   \hline
\rowcolor[HTML]{fdf3e7}
48  & Javed et al. \cite{JAVED2021102104} & 2021 & Bladder, Stomach, Colorectal adenocarcinomas, Breast, Kidney, Liver, Prostate
& 100 H\&E stained CRC histology, CoNSeP, PanNuke & Correlation filter
& 85\% Dice Score \\
\hline
  \rowcolor[HTML]{fdf3e7}
49  & Asif et al.\cite{correlation1} & 2017 & Large Intestine & 100 H\&E Colorectal adenocarcinomas 	& Correlation filter & 84\% F1Score  \\  
  \end{longtable}
 \end{tiny}
\paragraph{\textbf{ Adversarial Models based  Segmentation  }}
Generative Adversarial Networks (GANs) were first introduced by Good fellow et al. \cite{goodfellow2014generative} in 2014 with the basic idea of  generating synthetic images by mimicking the content of actual training datasets as described by Skandarani et al. \cite{Skandarani2021GANsFM}. 
With the introduction of GAN based models in several image processing tasks it was found that approaches relying on Generative Adversarial Networks have exhibited the capability of reducing the large annotated dataset requirements, thus reducing the potential barrier of automated image analysis in several medical imaging modalities as reviewed by Tschuchnig et al. \cite{gansurvey}. Table \ref{table7} summarize details about adversarial models based segmentation techniques. In the field of computational pathology recent GAN based developments have not only improved measures but have also enabled novel applications. Thus tasks relying on supervised techniques can now be performed via unsupervised techniques. In 2018, Zhang et al. \cite{gan2} proposed GAN based nuclei segmentation model as a dual contour enhanced adversarial network. In this approach contour highlighted and distance transformed masks are incorporated via adversarial network for improving cell nuclei segmentation. This approach outperformed previous state of the art models on MICCAI-2017 dataset, however this model was generalizable. Mahmood et al. \cite{gan4} proposed an improved model in 2020, via utilizing conditional GANs based training for nuclei segmentation on both real as well as synthetic data thus ensuring spatial consistency compared to conventional CNNs. A data augmentation based approach  proposed by Pandey et al. \cite{gan3} employed multi-GANs for improving performance of conventional segmentation approaches, one for generating synthesized mask which is incorporated into second GAN for performing conditional generation of synthesized image. Han et al. \cite {gan6} focused on unsupervised learning approach via generating 3D data using Recycle-GAN along with Hausdorff distance loss for nuclei shape preservation. For segmentation and classification of synthetically generated data a 3D CNN is employed. Yao et al. \cite{gan1} generated Scafold-A549 a very first synthetically generated nucleus segmentation dataset for training on varying density nuclei in 3D cell culture via recycle-GAN approach. Xing et al. \cite{gan5} proposed an adversarial two-directional domain adaptive method for nuclei detection on multiple modalities.
 Specifically, this method learns via a deep regression model through source to target and target to source image translation for each nuclei.


\begin{tiny}
\begin{longtable}{| m{0.4cm}| m{1cm} | m{1.6cm} | m{2.3cm} | m{1.8cm} | m{2cm} | m{1.8cm}|}
 
  \caption{GAN based Segmentation Methods} \label{table7} \\
   
  \hline
    \rowcolor[HTML]{FBEBD3}
\textbf{S.No}&
\textbf{Year}&
\textbf{Reference}&
\textbf{Organ}&
\textbf{Dataset}&
\textbf{Technique}& 
\textbf{Performance} \\ 
\hline
\endfirsthead

\multicolumn{3}{c}%
{{ }} \\
\hline 

\endhead

\hline 

\endfoot

\hline 
\endlastfoot
  
 \rowcolor[HTML]{fdf3e7}
       1 & 2018 & Zhang et al. \cite {gan2} & Head, Neck, Squamous cell, tumors & MICCAI-2017 & GAN  & 70\% F1 Score \\
        \hline
        \rowcolor[HTML]{fdf3e7}
       2 & 2020 & Pandey et al. \cite {gan3}  & Bladder, Colon, Stomach, Breast, Kidney, Liver, Pancreas, Colorectal & DSB-2018 & Multi-GAN  & 82\% Dice score  \\
        \hline
        \rowcolor[HTML]{fdf3e7}
       3 & 2020 & Mahmood et al. \cite {gan4} & Bladder, Colon, Stomach, Breast, Kidney, Liver & TCGA & GAN & 86\% F1 Score \\
        \hline
        \rowcolor[HTML]{fdf3e7}
        4 & 2020 & Han et al. \cite {gan6} & kidney & kidney dataset & Recycle-GAN   & 82\% F1 Score \\
        \hline
        \rowcolor[HTML]{fdf3e7}
       5 &  2021 & Yao et al. \cite {gan1} & Lungs & 3D Fluorescence image data & Cycle GAN & 50\% Dice Score\\
        \hline
        \rowcolor[HTML]{fdf3e7}
       6 & 2021 & Xing et al. \cite {gan5} & Colon & DAPI, TMI & Cycle GAN & 71\% F1 Score  \\
        \hline
   
  \end{longtable}
 \end{tiny}

\paragraph{\textbf{ Attention based Sequential Models }}
Attention mechanisms can be regarded as one of the hottest areas of deep learning research since last few years, originating primarily for natural language processing and now yielding excellent results in computer vision domain as well. They  works exactly on the principle of human eye, while viewing a scene in form of partial glimpses and paying enhanced attention to parts relevant to the context. In this way it not only focuses on selected regions but also concludes object interpretations at that particular point thus improving visual structure understanding. It explores global contextual information via building associations amongst features using attention mechanism, alongside adaptively aggregating long range contextual details, thus improving feature learning for accurate object segmentation. It can be considered as a tool for fair divisioning of allocated resources according to the quantity of information carried by signal. In most models its often used on top of higher level contextual information representing layer for better adaption amongst objects. Table \ref{table8} summarize details about attention based nuclei instance sequential models technique.

For medical image segmentation, first specific attention based model for nuclei segmentation task has been proposed in 2019 by Zhang et al. \cite{attention7}. They designed a binary tree network with two path fusion attention module via concatenating both low and high feature information followed by convolution for generating fused feature similar to a binary tree structure. A joint attention model based on Neural Architecture Spatial and channel weighting effect is proposed by Liu et al. \cite{attention2} using NAS search strategy for attention module automation with the addition of multiple attention module architectures searching within same network. A self supervised attention based nuclei segmentation approach is devised by Sahasrabudhe et al. \cite{attention5} from the assumption that nuclei texture and size could yield slide magnification thus generating self-supervised signal points for nuclei localization. 
For resolving prior noise issues a weakly supervised learning model is proposed by Guo et al. \cite{attention8} using nuclear centroid annotations for segmentation via generating  boundaries and super pixel masks as ground truth labels.
Salvi et al. \cite{watershedlatest} worked on weakly supervised learning approach for mitigating high quality annotated datasets requirement for training. They trained nuclei segmentation module via nuclei centroid annotation which were used for generating boundary and masks as ground truth label for segmentation and further performance enhancement is done through mask guided attention auxiliary network.

 For further refinement in previous approaches an encoder decoder based architecture and spatial channel joint attention module for abnormal nuclei detection is proposed by Ma et al. \cite{attention4} via using attention based merging technique for merging extracted varying and fixed features generated via R-CNN and fixed proposal module. Efficient feature extraction based approach is focused by Vahadane et al. \cite{attention6} and Lal et al. \cite{attention1} using 3 block architecture including residual bottleneck and attention decoder blocks. Robust residual blocks yielded high level object semantic maps while object localization is performed via attention module, thus improving accuracy. Zhao et al. \cite{9656603} employed a post processing pipeline for final segmentation from attention segmented image. They used a combination of morphological operations (binary  opening and closing) for coarse and fine object enhancements followed by distance transform and Gaussian blurring for local maxima identification and in the end watershed is used for final results.
\begin{tiny}
\begin{longtable}{| m{0.4cm}| m{1.2cm} | m{0.5cm} | m{2.2cm} | m{2.2cm} | m{2.5cm} | m{1.8cm}|}
 
  \caption{Attention based Nuclei Segmentation} \label{table8} \\
   
  \hline
    \rowcolor[HTML]{FBEBD3}
\textbf{S.No}&
\textbf{Reference}&
\textbf{Year}&
\textbf{Organ}&
\textbf{Dataset}&
\textbf{Technique}& 
\textbf{Performance} \\ 
\hline
\endfirsthead

\multicolumn{3}{c}%
{{ }} \\
\hline 

\endhead

\hline 

\endfoot

\hline 
\endlastfoot
 \rowcolor[HTML]{fdf3e7}
    1 & Lal et al.\cite{attention1} & 2021 & Liver, multiple & KMC Liver, Kumar Dataset	& Attention Mechanism, Encoder Decoder  & 70\% Dice Score  \\ 
    \hline
     \rowcolor[HTML]{fdf3e7}
    2 & Liu et al.\cite{attention2} & 2020 & Bladder, Colon, Stomach, Breast,  Kidney, Liver, Prostate & MoNuSAC &  Neural architecture based Spatial \& Channel Joint Attention Module & 86\% F1 Score \\
    \hline
     \rowcolor[HTML]{fdf3e7}
    3 & Aatresh et al.\cite{attention3} & 2021 & Kidney, Breast  & TNBC	& SegNet based attention module & 92\% F1 Score \\
    \hline
     \rowcolor[HTML]{fdf3e7}
    4 & Ma et al.\cite{attention4} & 2020 & Cervix & Herlev Dataset	& SE-FPM based  Self-attention & 92\% F1 Score \\
    \hline
     \rowcolor[HTML]{fdf3e7}
    5 & Sahasrabudhe et al.\cite{attention5} & 2020 & Bladder, Colon, Stomach, Breast, Kidney, Liver, Prostate & MoNuSeg & Self-Supervised attention module & 86\% F1 Score, 0.53 AJI \\
    \hline
     \rowcolor[HTML]{fdf3e7}
     6 & Vahadane et al.\cite{attention6} & 2021 & Bladder, Colon, Stomach, Breast, Kidney, Liver, Prostate & CPM-17, CoNSeP, Kumar &  U-Net, Attention, skip module & 81\% Dice Score \\
     \hline
      \rowcolor[HTML]{fdf3e7}
    7 & Zhang et al. \cite{attention7} & 2019 & Cervix & ISBI-2014 Dataset & ResNext, Two path fusion binary tree & 90\% F1 Score\\
    \hline
     \rowcolor[HTML]{fdf3e7}
    8 & Gunesli et al. \cite{gland9} & 2020 & Colon & Pathology Department Hacettepe University & FCN & 94\% F1 Score \\
    \hline
    
  \rowcolor[HTML]{fdf3e7}
    9 & Guo et al. \cite{attention8} & 2021 & Bladder, Colon, Stomach, Breast, Kidney, Liver, Prostate & MoNuSeg, TNBC & Mask guided attention network & 83\% F1 Score \\

    \hline
      \rowcolor[HTML]{fdf3e7}

    10 & Zhao et al. \cite{9656603} & 2021 & Bladder, Colon, Stomach, Breast, Kidney, Liver, Prostate & ISBI-2014 Dataset, BNS, MoNuSeg & U-NET, Attention & 72\% AJI\\
    \hline
   
  \end{longtable}
 \end{tiny}


\paragraph{\textbf{Stand out Method}}
\textbf{S}egNet based attention guided architecture proposed by Aatresh et al. \cite{attention3} and Mask guided attention network by Guo et al. \cite{attention8}  resulted in 92\% F1- Score and stood as top performing  attention based instance segmentation techniques.

\subsection{ \textbf{Glands Instance Segmentation Methods}}

\paragraph{\textbf{Handcrafted features extraction}}
For medical image analysis, traditional handcrafted feature based techniques are more prevalent for segmentation compared to learning based approaches. Table \ref{table9} summarize details about features extraction techniques for glands instance segmentation. Classical methods depends heavily on image features including its color, shape and texture primarily. Similarly, for instance segmentation of natural images mostly pipeline comprises of object detection and masking. In 2017, Zarei et al. \cite{gland1} proposed a gland segmentation pipeline via integrating classical techniques. Firstly, digitized histopathological images are constructed using sixteen light wavelengths followed by RGB construction by Principal Component Analysis (PCA). For glands segmentation unsupervised clustering is applied and utilized morphological cleaning operation for small objects removal and then eroded processed gray scale image for further segmentation. Despite being novelty of this technique, generalizability and dataset size variation was the major loop hole.
 Similarly, Wang et al. \cite{gland3}  used contour based energy minimization technique for efficiently segmenting glands instances from background.
Two level sets are utilized including energy based and region based models. Model divided image into 2 parts, glands
with lumens and glands without lumens. Glands with lumens were localized and segmented via edge-based level set. Similarly, glands without lumens stromal were segmented via region based technique. For output prediction results of both these models are combined together.
 Manivannan et al. \cite{gland6} combined both traditional handcrafted multi scale features, with features learned through deep convolutional network trained for mapping images to respective segmentation output.They used structured learning approach for capturing structural information  of image e.g (location between glands and neighbouring glands identification as separate instance). These were then used for training support vector machine classifier and is further combined and post processed for segmentation output. Rezai et al. \cite{gland7} preprocessed input data via invariant local binary pattern  based classical features processing, extracted features were further passed to LinkNet for glands segmentation training.

\paragraph{\textbf{ Deep learning based Glands Instance Segmentation }}

In 2017 Xu et al. \cite{gland2} proposed neural networks glands segmentation algorithm based on image-image prediction model using deep multi channel model for automatic fusion of complex multi-channel information based on regional,
local and boundary pixel features. This model reduced heavy feature designs issue due to the use of CNNs. Similarly, alternate channels resulted in better feature learning resulting in better performance for training single scale objects. However, problem with multi stage learning was still there. For incorporating multi scale objects segmentation a minimal information dilated network is proposed by Graham et al. \cite{gland8} where for segmenting varying sized glands, maximal information is retained during feature extraction on which atrous spatial pyramid pooling is applied. Final level evaluation of keeping or discarding predictions is done through object level uncertainty score. Considering the computational cost issues in prior approaches a cost efficient and adaptive multi stage attention based learning module is designed by Gunesli et al. \cite{gland9} for adaptively learning hard to learn pixels at each stage on given image data without any prior preprocessing via multi stage boosting network for in parallel adaptive learning and  pixels correction at the same time by adjusting loss weight for each predicted pixel. A unique boundary adjustment loss function is used for paying  focused attention to pixels nearer to boundaries.

 Previously designed glands segmentation methods used several deep learning features and auxiliary contour prediction output maps for modeling segmentation tasks. However, they fail to capture complex structural variations in gland images thus resulting in  contour imbalance problem due to limited performance. To overcome these problems, Mei et al. \cite{gland11} proposed a dense contour-imbalance aware (DCIA) framework including convolutional neural network (DenseNet) and focal loss (FL). In this technique features generated via DenseNet were explored for “optimal” image representation and focal loss function mitigated the contour imbalance problem in the training stage. Finally, for fine tuning predicted confidence maps, post-processing is done via morphological operations and convolutional conditional random fields (ConvCRFs). A single deep learning shape adversarial domain adaptation model for accurate segmentation of glands is proposed by Yan et al. \cite{gland12} where a segment level shape similarity measure is used for curve similarity calculation between each annotated boundary and corresponding boundary segment detection, images down sampled at multiple scales were integrated for context enabled global as well as local features training. 
  Ding et al. \cite{gland13} proposed a multi scale fully convolutional network and three class classification (TCC-MSFCN) framework for better segmentation approximation. Multi scale architecture extracted varying receptive field features corresponding to object size. Similarly, for global information loss computation, separate high-resolution branch is included in model. Finally, for accurate segmentation of touching glands, three-class classification based on edge pixels is applied.
  Rastogi et al. \cite{glandnca} proposed an encoder decoder based module for better capturing of contextual information via harnessing multi level CNN capabilities as well as features concatenation and features upsampling via skip connections and harnessed the exceptional power of neural networks for capturing contextual information and features concatenation via upsampling. For fine tuning raw predicted samples are processed using morphological opening and closing operators yielding 85\% accuracy.
Xie et al. \cite{10.1007/978-3-030-59722-1_40} proposed a pairwise relations learning module for enhancing image representation ability by exploiting the semantic consistency between image pairs and transferring learned features to S-Net an encoder decoder based network for improving segmentation.
Salvi et al. \cite{salvi} proposed a rapid gland identification module for prostate gland segmentation via similar multi-channel algorithm. However, in this approach both traditional and deep learning techniques are exploited and fused together via a hybrid instance segmentation strategy based on stroma detection for accurate detection and delineation of target gland contours.
Dabass et al. \cite{gland16}  with a slight enhancement of dimension wise convolution combined with atrous spatial pyramid pooling for improved encoder decoder architecture and multi task learning based on region and boundaries extraction along with feature learning, diminishing gradient issue is alleviated as well. Similarly,  Wen et al. \cite{gland17} proposed Gabor encoder based module for texture enabled feature learning through cascaded squeeze parsing to Bi-Attention mechanism for capturing both channel and spatial information at multiple scales followed by class balancing.

\paragraph{\textbf{Stand Out Method}}
In comparison to segmentation results obtained through handcrafted segmentation techniques, we found that FCN based Atrous spatial pyramid pooling technique by Graham et al. \cite{gland8} outperformed all previous state of the art proposed minimally tuned classical algorithms through a large margin, producing better segmentation outputs over all coverage thresholds. 
\begin{tiny}
\begin{longtable}{| m{0.4cm}| m{1.7cm} | m{0.8cm} | m{1.2cm} | m{2cm} | m{3cm} | m{1.5cm}|}
 
  \caption{Glands Instance Segmentation Techniques Summary} \label{table9}\\
   
  \hline
    \rowcolor[HTML]{FBEBD3}
    \textbf{S.No} &
    \textbf{Reference}&
    \textbf{Year}&
    \textbf{Organ}&
    \textbf{Dataset}&
    \textbf{Technique}& 
    \textbf{Performance} \\ 
\hline
\endfirsthead

\multicolumn{3}{c}%
{{ }} \\
\hline 

\endhead

\hline 

\endfoot

\hline 
\endlastfoot

\rowcolor[HTML]{fdf3e7}
1  & Zarei et al.\cite{gland1} & 2017 & Prostate & TMA	& PCA, Clustering, Morphological Operation & 80\% Dice Score \\  
\hline
\rowcolor[HTML]{fdf3e7}
2 & Xu et al.\cite{gland2} & 2017 & Colon & Warwick-QU	& 3-channel Conv-Net, FCN, HED edge detection & 83\% F1Score \\

\hline
 \rowcolor[HTML]{fdf3e7}
3 &  Wang et al.\cite{gland3} & 2017 & Endometrial & West China Hospital	& Level set edge based energy minimization \& region & 75\% Dice Score \\   

\hline
\rowcolor[HTML]{fdf3e7}
4 & Wenqi et al.\cite{gland6} & 2018 & Colon & Warwick-QU & FCN \& SVM classifier & 89\% Dice Score \\
\hline
\rowcolor[HTML]{fdf3e7}
5 & Rezai et al.\cite{gland7} & 2019 & Colon & Warwick-QU & FCN \& LinkNet \& Local binary pattern & 82\% Dice Score \\
\hline
\rowcolor[HTML]{fdf3e7}
\textbf{6} & \textbf{Graham et al.}\cite{gland8} & \textbf{2019} & \textbf{Colon} & \textbf{GlaS} & \textbf{FCN} \& \textbf{Atrous Spatial Pyramid Pooling} & \textbf{94\% F1 Score} \\
\hline
\rowcolor[HTML]{fdf3e7}
7 & Gunesli et al.\cite{gland9} & 2020 & Prostate & Pathology Dept.	Hacettepe University & Attention Boost, Iterative Multistage Learning & 95\% F1 Score \\ 
\hline
\rowcolor[HTML]{fdf3e7}
8 & Mei et al.\cite{gland11} & 2020 & Colon & Warwick-QU &  CRF, Morphological Operations, CNN & 79\% F1 Score \\
\hline
\rowcolor[HTML]{fdf3e7}
9 &  Yan et al.\cite{gland12} & 2020 & Liver bone & CRAG &  Weighted Matrix Adversarial Loss & NA \\
\hline
\rowcolor[HTML]{fdf3e7}
10  & Ding et al.\cite{gland13} & 2020 & Colon, Liver & CRAG Warwick-QU & TCC-MSFCN Network & 91\% F1 Score \\
\hline
\rowcolor[HTML]{fdf3e7}
11 & Yutong et al.\cite{10.1007/978-3-030-59722-1_40} & 2020 & Colon & CRAG, GlaS & Pairwise Relational Module, S-Net & 87\% Dice Score \\
\hline
\rowcolor[HTML]{fdf3e7}
12 & Rastogi et al. \cite{glandnca} & 2021 & Colon & Warwick-QU & U-Net & 85\% F1  \\
\hline
\rowcolor[HTML]{fdf3e7}
13 &  Wen et al. \cite{gland17}   & 2021 & Colon, Prostate  & GlaS dataset, CRAG & Gabor Filter, Attention Module & 83\% F1 Score \\
\hline
\rowcolor[HTML]{fdf3e7}
14 &  Salvi et al. \cite{salvi}   & 2021 & Prostate gland & GlaS dataset & RINGS, ACM  & 90\% Dice Score\\
\hline
\rowcolor[HTML]{fdf3e7}
15 & Dabass et al.\cite{gland16} & 2021 & Colon & GlaS, CRAG & U-Net, Attention Mechanism & 93\% F1 Score  \\
  \end{longtable}
 \end{tiny}

\section {Discussion \& Future Prospects}

This is the first study that reviewed both nuclei and glands instance segmentation techniques evolved during last 5 years for multiple organs included but not limited to  liver, kidney, prostate, bladder, colon, stomach, lung, and brain cancer. 
This survey covers both handcrafted features extraction methods as well as deep convolutional neural networks based techniques emerged during this time span. Overall, This has been observed that deep learning methods outperformed all traditional segmentation methods. Best performing model was attention based U-Net techniques covering both local as well as global features mapping and multi scale refinement. Region proposal based Mask- RCNN accuracy also reported upto the mark accuracy with different optimization parameters and loss functions. For MoNuSeg dataset,  Mask-RCNN models performance was found even better than U-Net. Transfer learning based feature pre-training also yielded better results and model trained independent of tissue type for nuclei and glands instance segmentation respectively. This review summarises major evolution and advancements adopted by researchers in model designing for the task of instance segmentation. Overall, In histopathology whole slide image analysis, base techniques provide novel architectures for 
clinical workflows automation via automatic WSI feature extraction that serves a pivotal role in diagnosis, treatment, and survival prediction of various lethal diseases.
Till date, advancements in digital pathology has automated complete histopathological cancer grading process, mitosis detection, cancer sub-typing, tumor classification \& segmentation. 

This all has been made possible via deep learning based networks that enabled training possible for large scale highly varying whole slide images at contrasting magnifications.  Rapid advancements in the field of  oncology will lead novel innovations and insights  in tumor nuclei instance segmentation and feature extraction thus yielding better cancer treatment selection methods. One such way would be AI driven diagnostic techniques leveraging deep leaning architectures. 

Encoder Decoder based U-Net techniques gave excellent results for biomedical images segmentation. Despite its great performance model struggles during classification of closely touching instances. Symmetrical network architectures further possess an opportunity to modify model structure and improve performance accuracy. Since, initial U-Net architecture followed a typical CNN based structure, having repetitive convolution, activation, and pooling layers for feature maps calculation. 

However, with the rising complexity of data, training DCNNs resulted in further network architecture advancement and leads towards other novel networks including HoVer-Net \cite{graham2019hover}, SegNet \cite{attention3}, ResUNet \cite{resunet1}, ResNet \cite{resu2}, PR-Net \cite{10.1007/978-3-030-59722-1_40}, ResNext \cite{attention7}, OS-Net \cite{UNET1} and various others. These comparatively new approaches have embarked excellent state-of-the-art instance segmentation results with better feature extraction thus minimizing vanishing gradient problem, and improving network convergence. Performance boost primarily resulted by modification from base network 
architecture to modified architectures, either through replacement of initial skip connection layers with longer ones or by layers addition in base U-Net model for deeper neural network. Symmetrical network architectures further possess an opportunity to modify initial network structure and improve performance accuracy.

\begin{figure}[H]%
\centering
\includegraphics[width=1\textwidth]{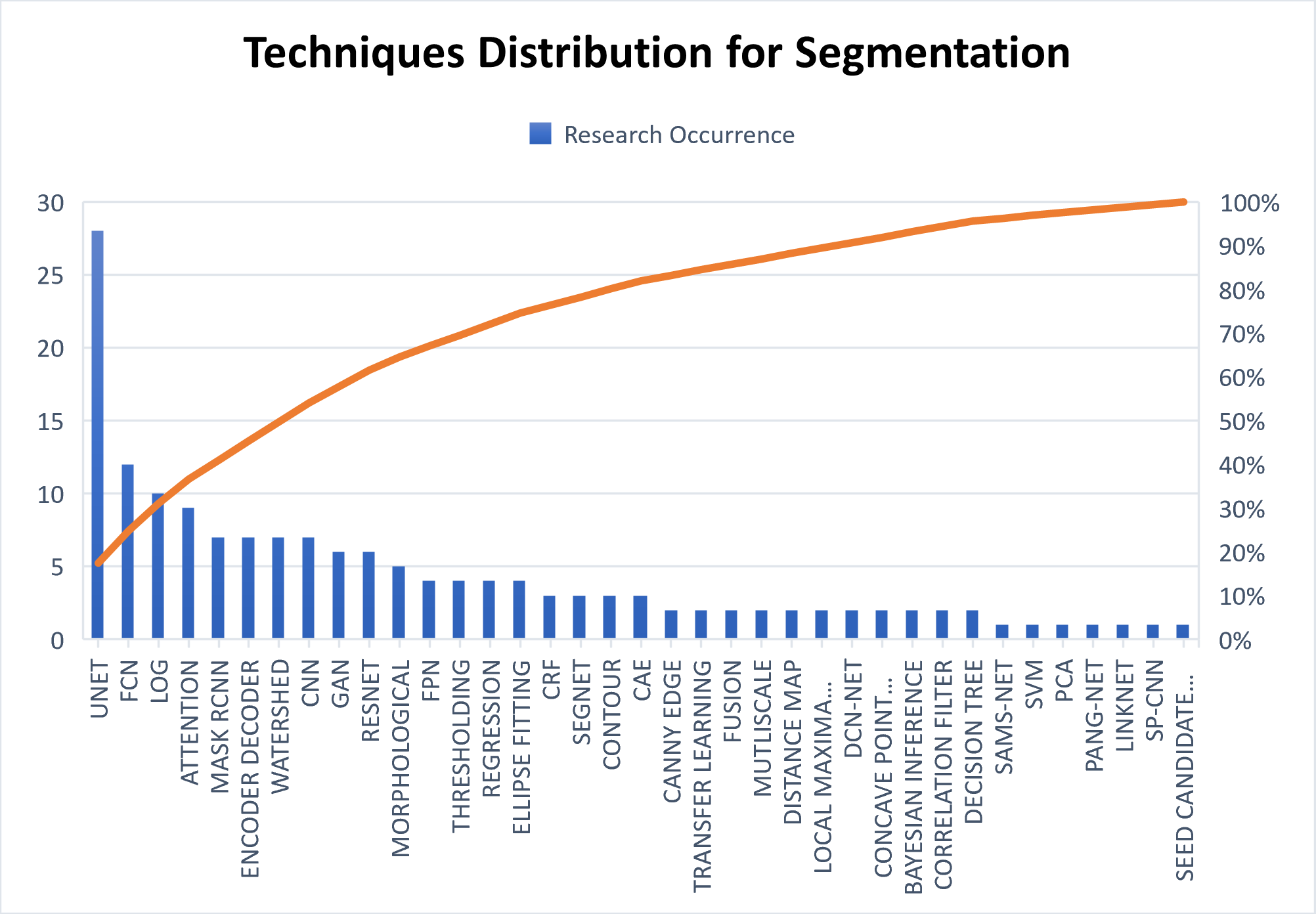}
\caption{Sum of Research Occurrence for each Tool.  Color shows details about Tool.  Size shows sum of Research Occurrence. The view is filtered on sum of Research Occurrence, which ranges from 1 to 28.}\label{fig14}
\end{figure}

Figure \ref{fig14} shows cumulative techniques occurrence in research articles  reviewed for this paper.

Multistage learning architectures have attained huge success in domain of medical imaging specially segmentation tasks. However information loss at middle layers along with inconsistent features learning is the major common drawback of these architectures. Another primary limitations observed is often times they leads to a redundant information usage when similar low level feature are repeatedly extracted several times at multiple scales.

Similarly, whole slide images are usually regarded as texture images while fixed 
encoder decoder based architecture of U-Net doesnot serve as an appropriate model for texture based learning models because of its inability to extract features at different scales and orientations.
 It has been observed that malignant tumors have a particularly higher growth rate compared to benign and greater density as well causing nuclei overlapping or occlusion during squeezing operation.
One important challenge is sufficient networks training to yield good generalizations for hard-to-learn pixels. A typical group of such hard-to-learn pixels are boundaries between instances.
In spite of the huge success of FCNs trained on very large datasets,
training becomes difficult when small quantities of annotated data are available and when pixels of background and foreground classes are highly imbalanced. In such cases, without further adjustments, network tend to yield poor generalizations for pixels of a minority class as well as for hard-to-learn pixels. Similarly, in some approaches for long-range, features dependencies were not modeled efficiently thus leading to non-optimal discriminative representation of features.

Two stage learning based Mask-RCNN network is proposed for instance level annotations generation. In first stage it proposes class probability maps while its CNN backbone provides input feature map which is then fed in both regression as well as classification layer. Regression layer predicts region proposals while classification layer predicts object existence probability within the region proposal. Similarly, second stage of Mask-RCNN utilized generated region proposals for object classification, bounding box prediction and final segmentation.

Region of interest (RoI) alignment was introduced with Mask-RCNN.
It incorporated bilinear interpolation for floating point values calculation from sampling points. It resulted in reduced computational time, by taking RoI defined feature maps and its scaling to fixed sized patches. Mask-RCNN variants output bounding box coordinates, objects classification and segmentation map for each object instance.

 Although, R-CNN, Faster-RCNN and its other variants have achieved good performance in general target detection tasks by combining semantic segmentation and target detection, but their performance in histopathology tasks such as abnormal cell detection was still not as per expectations causing issues like over and under segmentation which later yielded incorrect estimation of the nuclear density, size and morphology. Another primary limitation of the region proposal based methods is their difficulty in merging instances prediction while processing neighbouring tiles e.g in a condition when a nuclei sub segment at boundary is assigned a label this thing needs to be ensured that rest of the nuclei part present in neighbouring patch should be assigned same label.
 
 Similarly, deep learning models also lacks theoretical understanding thus leading to degraded performance when limited training data is provided.
 
Pre-deep learning techniques performed exceptionally well for small amount of data  where training and testing inputs were taken from single feature space. Handcrafted segmentation techniques primarily used in majority of the approaches were intensity based thresholding, morphological operations including erosion \& dilation, marker controlled watershed transform, active contour models, with machine learning classifiers such as RVM \cite{Tipping1999TheRV}, SVM \cite{10.1007/978-3-642-34041-3_27}, KNN \cite{10.1007/978-3-540-39964-3_62} clustering and other supervised classification methods.

 Thresholding and morphological transformations are usually regarded as basic image processing operations but for nuclei having complex image background, segmentation via these operations gets difficult since thresholding techniques such as Otsu usually works with the assumption of nuclei having distinct intensity and thus fails in most cases where tumor cells display high level of variability in chromatin appearance. Watershed  performance was fine given target locations but in all other scenarios shows poor performance, since it considers a relatively homogeneous nuclei or gland appearance and a pre requisite of initially detected seed points before segmentation.

 Active contour models or level Sets deals with change in spatial temporal relations (i.e combination of image and shape features inside nuclei) but fails in handling boundary events and alike watershed this technique is also dependent on initialization of good seed points. Thus despite being in use for long, there are still problems in segmentation results of these techniques because they are based on simple assumption about the nucleus appearance characterstics including low intensity or circular borders which may not be true in all cases (e.g. some nuclei exhibits high intensity and irregular shapes \cite{7950548} thus yielding low accuracy or poor precision. 
  Similarly  clustering and graph based methods have been used too, but their computationally expensive nature and shallow feature learning consideration makes them a misfit for training large scale data \cite{8363645}.
Similarly, Radiating gradient vector flow (RGVF) snake and graph cuts techniques were used for fine segmentation but these do not involve inherent shape constraints for the segmented boundary, thus resulting in irregular nucleus boundary.

For sparse data conditions, transfer learning techniques can be utilized using pre-trained weights of a comparatively larger model for training similar smaller dataset networks given the domain similarity checks.   

 A training limitation in majority of datasets is small amount of available data for learning, validation and further testing due to the limited number of open source datasets. Similarly, data collection and preprocessing stages during samples collections such as manual H\&E staining or varying slide scanners at times results in color inconsistency \cite{challenge1} and unclear nuclear boundaries and this drawback later effects entire model training and testing stages.
 Future work involves expanding tissue types and number of images included in the dataset. Adversarial learning based generative adversarial networks (GANs) serves as dual learning techniques that first generates synthetic whole slide images followed by segmentation network for data training. GANs are considered as a promising way of data synthesis, that generates H\&E stained synthetic histopathological images yielding ground truths from learned features from training data.

\section {Conclusion}
Segmentation of nuclei  and glands instances from histopathological images, is the focus of attention because of its vitality in biomedical and biological applications. Automated glands and nuclei segmentation is fundamental to chronic disease diagnosis, survival prediction, phenotype classification, feature extraction and cell tracking. It also has significance in cancer diagnosis, grading, and analysis as they are highly dependent on the quality of nuclei segmentation. Despite the significant advancement in automated segmentation, separation of large cluster of nuclei, irregular glands structure and outlining their boundaries with a high precision and speed is still considered as a challenge. Apart from that, researchers are still lacking a generalized benchmark solution for all types of nuclei and glands instance segmentation from distinct histopathological images under various condition. Due to emergence of Convolutional neural networks (CNNs) the classical hand-crafted feature extraction techniques are replaced by CNNs. This review article covers the most recent approaches of last five years for nuclei and glands instance segmentation, along with major publicly available datasets and summary of the grand challenges held specific to this task. The review compiled deep learning computational approaches, hand-crafted morphological feature-based approaches for evaluating advantages and draw backs of each segmented technique. The hindrances in the nuclei instance segmentation such as varying staining impacts, insufficient data causing over fitting, disparate nuclei and glands structure, models specificity to a single image set are discussed.\\

 \bibliographystyle{elsarticle-num} 
 \bibliography{cas-refs}





\end{document}